\documentclass[longauth]{aa}  
\usepackage{graphicx}
\usepackage{txfonts}
\usepackage[colorlinks=true, citecolor=blue]{hyperref}
\usepackage{multirow}
\usepackage{booktabs}
\usepackage{lscape}
\usepackage{placeins}

\newcommand{\ha}{H$\alpha$}
\newcommand{\lya}{Ly$\alpha$}
\newcommand{\oiii}{[O\,{\footnotesize III}]}

\newcommand{\nii}{[N\,{\footnotesize II}]}
\newcommand{\sii}{[S\,{\footnotesize II}]}
\newcommand{\kms}{km s$^{-1}$}
\newcommand{\myr}{$M_{\odot}$ yr$^{-1}$}
\newcommand{\msun}{$M_{\odot}$}
\newcommand{\zsun}{$Z_{\odot}$}
\newcommand{\mstar}{$M_{\star}$}
\newcommand{\mgas}{$M_{\rm gas}$}

\newcommand{\mdust}{$M_{\rm dust}$}

\newcommand{\kkmspc}{K~km~s$^{-1}$~pc$^2$}

\newcommand{\jykms}{Jy~km~s$^{-1}$}
\newcommand{\ci}{[C\,{\footnotesize I}]}
\newcommand{\hi}{H\,{\footnotesize I}}
\newcommand{\hii}{H\,{\footnotesize II}}
\newcommand{\co}{CO}
\newcommand{\citwo}{[C\,{\footnotesize I}]$(^3P_2\,-\, ^{3}P_1)$}
\newcommand{\coseven}{CO\,$(7-6)$}

\newcommand{\lprimecitwo}{$L'_{\mathrm{[C\,\scriptscriptstyle{I}\scriptstyle{]}}^3P_2\,-\, ^3P_1}$}
\newcommand{\lprimecoseven}{$L'_{\rm CO(7-6)}$}
\newcommand{\lir}{$L_{\rm IR}$}

\newcommand{\lsun}{$L_{\odot}$}
\newcommand{\tdust}{$T_{\rm dust}$}

\newcommand{\subl}{sub-$L^{\star}$}
\newcommand{\cii}{[C\,{\footnotesize II}]}
\newcommand{\jwst}{JWST}
\newcommand{\hst}{HST}

\begin{document} 

   \title{The cold interstellar medium of a normal sub-$L^\star$ galaxy \\ at the end of reionization}

   \author{F. Valentino\inst{1,2},
          S. Fujimoto\inst{3},
          C. Gim\'{e}nez-Arteaga\inst{2,4},
          G. Brammer\inst{2,4},
          K. Kohno\inst{5,6},
          F. Sun\inst{7},
          V. Kokorev\inst{8},
          F. E. Bauer\inst{9,10,11,12}, 
          C. Di Cesare\inst{13,14,15},
          D. Espada\inst{16,17},
          M. Lee\inst{2,18},  
          M. Dessauges-Zavadsky\inst{19},
          Y. Ao\inst{20,21},
          A. M. Koekemoer\inst{22},
          M. Ouchi\inst{23,24,25},
          J. F. Wu\inst{22,26},
          E. Egami\inst{7},
          J.-B. Jolly\inst{27},
          C. del P. Lagos\inst{28,29,2},
          G. E. Magdis\inst{2,4,18},
          D. Schaerer\inst{19},
          K. Shimasaku\inst{5,6},
          H. Umehata\inst{30,31,32},
          W.-H. Wang\inst{33}
          }
    \institute{European Southern Observatory, Karl-Schwarzschild-Str. 2, 85748 Garching, Germany
    \and Cosmic Dawn Center (DAWN), Denmark
    \and Department of Astronomy, The University of Texas at Austin, Austin, TX, USA
    \and Niels Bohr Institute, University of Copenhagen, Jagtvej 128, DK-2200 Copenhagen N, Denmark
    \and Institute of Astronomy, Graduate School of Science, The University of Tokyo, 2-21-1, Osawa, Mitaka, Tokyo 181-0015, Japan
    \and Research Center for Early Universe, Graduate School of Science, The University of Tokyo, 7-3-1, Hongo, Bunkyo-ku, Tokyo 113-0033, Japan 
    \and Steward Observatory, University of Arizona, 933 N. Cherry Ave., Tucson, AZ 85721, USA
    \and Kapteyn Astronomical Institute, University of Groningen, 9700 AV Groningen, The Netherlands
    \and Instituto de Astrof{\'{\i}}sica, Facultad de F{\'{i}}sica, Pontificia Universidad Cat{\'{o}}lica de Chile, Campus San Joaquín, Av. Vicuña Mackenna 4860, Macul Santiago, Chile, 7820436
    \and Centro de Astroingenier{\'{\i}}a, Facultad de F{\'{i}}sica, Pontificia Universidad Cat{\'{o}}lica de Chile, Campus San Joaquín, Av. Vicuña Mackenna 4860, Macul Santiago, Chile, 7820436
    \and Millennium Institute of Astrophysics, Nuncio Monse{\~{n}}or S{\'{o}}tero Sanz 100, Of 104, Providencia, Santiago, Chile
    \and Space Science Institute, 4750 Walnut Street, Suite 205, Boulder, Colorado 80301
    \and Dipartimento di Fisica, Sapienza, Universita` di Roma, Piazzale Aldo Moro 5, I-00185 Roma, Italy
    \and INFN, Sezione di Roma I, Piazzale Aldo Moro 2, I-00185 Roma, Italy
    \and INAF/Osservatorio Astronomico di Roma, Via di Frascati 33, I-00078 Monte Porzio Catone, Italy
    \and Departamento de F\'{i}sica Te\'{o}rica y del Cosmos, Campus de Fuentenueva,
Edificio Mecenas, Universidad de Granada, E-18071 Granada, Spain
    \and Instituto Carlos I de F\'{i}sica Te\'{o}rica y Computacional, Facultad de Ciencias, E-18071 Granada, Spain
    \and DTU-Space, Technical University of Denmark, Elektrovej 327, DK-2800 Kgs. Lyngby, Denmark
    \and Department of Astronomy, University of Geneva, Chemin Pegasi 51, 1290 Versoix, Switzerland
    \and Purple Mountain Observatory and Key Laboratory for Radio Astronomy, Chinese Academy of Sciences, Nanjing, PR China
    \and School of Astronomy and Space Science, University of Science and Technology of China, Hefei, PR China
    \and Space Telescope Science Institute, 3700 San Martin Dr., Baltimore, MD 21218, USA
    \and Institute for Cosmic Ray Research, The University of Tokyo, 5-1-5 Kashiwanoha, Kashiwa, Chiba 277-8582, Japan
    \and Kavli IPMU (WPI), The University of Tokyo, 5-1-5 Kashiwanoha, Kashiwa, Chiba 277-8583, Japan
    \and National Astronomical Observatory of Japan, 2-21-1 Osawa, Mitaka, Tokyo 181-8588, Japan
    \and Department of Physics \& Astronomy, Johns Hopkins University, 3400 North Charles Street, Baltimore, MD 21218, USA
    \and Max-Planck-Institut für Extraterrestrische Physik (MPE), Giessenbachstraße 1, D-85748 Garching, Germany
    \and ARC Centre of Excellence for All Sky Astrophysics in 3 Dimensions (ASTRO 3D), Australia
    \and International Centre for Radio Astronomy Research (ICRAR), University of Western Australia, Crawley, WA 6009, Australia
    \and Institute for Advanced Research, Nagoya University, Furocho, Chikusa, Nagoya 464-8602, Japan 
    \and Department of Physics, Graduate School of Science, Nagoya University, Furocho, Chikusa, Nagoya 464-8602, Japan 
    \and Cahill Center for Astronomy and Astrophysics, California Institute of Technology, MS 249-17, Pasadena, CA 91125, USA
    \and Academia Sinica Institute of Astronomy and Astrophysics (ASIAA), No. 1, Sec. 4, Roosevelt Rd., Taipei 10617, Taiwan
    }
       
  \authorrunning{Valentino, Fujimoto, et al.}
  \titlerunning{Molecular gas and dust in a normal sub-$L^\star$ galaxy at $z=6$}
   \date{Received --; accepted --}

  \abstract{We present the results of a $\sim$60-hour multiband observational campaign with the Atacama Large Millimeter Array targeting a spectroscopically confirmed and lensed \subl\ galaxy at $z=6.07$, first identified during the ALMA Lensing Cluster Survey (ALCS). We sampled the dust continuum emission from rest frame 90 to 370 $\mu$m at six different frequencies and set constraining upper limits on the molecular gas line emission and content by targeting the \coseven\ and \citwo\ transitions in two lensed images with $\mu\gtrsim20$. Complementing these submillimeter observations with deep optical and near-IR photometry and spectroscopy with \jwst, we find this galaxy to form stars at a rate of $\mathrm{SFR} \sim7$ \myr, $\sim 50-70$\% of which is obscured by dust. This is consistent with what one would predict for a $M_\star\sim 7.5\times10^{8}$ \msun\ object by extrapolating the relation between the fraction of the obscured star formation rate and stellar mass at $z<2.5$ and with observations of IR-detected objects at $5<z<7$. The light-weighted dust temperature of $T_{\rm dust}\sim50$ K is similar to that of more massive galaxies at similar redshifts, although with large uncertainties and with possible negative gradients. We measure a dust mass of $M_{\rm dust} \sim 1.5 \times 
 10^6 \, M_\odot$ and, by combining \ci, \cii, and a dynamical estimate, a gas mass of $M_{\rm gas}\sim2\times10^9$ \msun. Their ratio ($\delta_{\rm DGR}$) is in good agreement with predictions from models and empirical relations in the literature.
 The dust-to-stellar mass fraction of $f_{\rm dust}\sim 0.002$ and the young stellar age ($100-200$ Myr) are consistent with efficient dust production via supernovae, as predicted by existing models and simulations of dust evolution. Also, the expected number density of galaxies with \mdust\ $\sim10^{6}$ \msun\ at $z=6$ from a subset of these models is in agreement with the observational estimate that we set from the parent ALCS survey. The combination of gravitational lensing and deep multiwavelength observations allowed us to probe luminosity and mass regimes up to two orders of magnitude lower than what has been explored so far for field galaxies at similar redshifts. Our results serve as a benchmark for future observational endeavors of the high-redshift and faint \subl\ galaxy population that might have driven the reionization of the Universe.}  
   \maketitle
%
\section{Introduction}
Over the past few years, the number of galaxies confirmed at redshift $z>6$ and deeper into the reionization epoch has soared. Interferometric (sub)millimeter observations and ground-based and, more recently, space-based optical and near-IR spectroscopy have been instrumental in allowing us to start exploring the physics regulating the growth of the first galaxies. Truly multiwavelength studies proved to be necessary to observe and connect all components in galaxies and provide a complete view of these systems. 
However, for reasons of opportunity and observing time cost, priority has been given to the brightest and rarest targets that could
maximize the detection rates -- but in doing so we have been missing more numerous, typical
galaxy populations at $z>6$.\\

In light of the possible
preponderant role played by average, rather than exceptional, galaxy populations in the reionization of the
Universe \citep{robertson_2022}, there has been a renewed focus on spectroscopic studies of faint sources around or below the knee of
the luminosity function ($L^\star$) and on the main sequence of star formation \citep{daddi_2007} at
these redshifts, particularly after the launch of the \textit{James Webb} Space Telescope (\jwst). Moreover, observational campaigns of faint objects are even more affordable when the Universe comes to our aid with the
gravitational lensing effect. This phenomenon provides a unique window onto the formation of faint galaxies on small scales, which would otherwise be impossible to probe without the presence of massive objects along the line of sight.

\begin{table*}[]
    \centering
    \caption{Summary of the ALMA observations presented here.}
    \begin{tabular}{lcccccc}
    \toprule
    \toprule
       Band & $t_{\rm int}$ & Frequency&  $\lambda_{\rm rest}$&  Continuum rms & Beam size& Program ID\\
            &   \small [hr]&    \small [GHz]&  \small [$\mu$m]& \small [$\mu$Jy/beam] & \small [$\arcsec\,\times\,\arcsec$]& \\ 
    \midrule
    Band 8 & 0.6  & 478.8 & 88  & 2458 & $0.65 \times 0.57$ & 1, 2 \\
    Band 7 & 2.0  & 345.9 & 122 & 183  & $0.93 \times 0.69$ & 1, 2 \\
    Band 7 & 1.1  & 293.0 & 146 & 149  & $0.84 \times 0.67$ & 3 \\
    Band 6 & 2.6  & 268.7 & 158 & 77   & $0.44 \times 0.38$ & 3 \\
    Band 5 & 1.8  & 205.0 & 205 & 67   & $1.08 \times 0.83$ & 1, 2 \\
    Band 3 & 13.3 & 100.5 & 370 & 41   & $1.55 \times 1.27$ & 4 \\
    \bottomrule
  \end{tabular}
  \tablefoot{\smallskip Program IDs: (1): 2022.1.00195.S (PI: S. Fujimoto); (2): 2021.1.00247.S (PI: S. Fujimoto); (3): 2021.1.00055.S (PI: S. Fujimoto); (4): 2021.1.00181.S (PI: F. Valentino). The on-source integration times $t_{\rm int}$ are for each z6.3 and arc image. Band 6 were obtained in C5 ($t_{\rm int} = 1.8$ hr on target per source, $0.35\arcsec \times 0.29\arcsec$ beam) and C2 ($t_{\rm int} = 0.8$ hr, $1.26\arcsec \times 1.05\arcsec$ beam) and combined with \textsc{CASA}.   
  The rms per beam in the continuum maps is estimated as described in Section \ref{sec:photometry}.}
    \label{tab:obs}
\end{table*}

Here we attempt to push the existing boundaries for deep multiwavelength extragalactic studies at high redshifts by leveraging state-of-the-art instruments and the lensing effect. 
Our primary objective is to start exploring a new portion of the cold gas and dust parameter space at intrinsic low stellar masses, star formation rates (SFRs), and metallicities. These regimes are critical to understanding the first phases of galaxy formation. The availability of cold gas reservoirs and the impact of intense, hard radiation feedback from young stars with low metallicities ultimately regulate the growth of the numerous population of low-mass galaxies and their ability to reionize the Universe. Furthermore, these processes are intricately linked to the rate at which metals and dust accumulate, and the relationship between these two components in the earliest phases of galaxy assembly remains itself a debated topic \citep[see, e.g.,][for a few recent examples]{peroux_2020, popping_2023, heintz_2023_dtg, konstantopoulou_2024, schneider_2023}. Fortunately, both dust and metals can now be traced in early galaxies. Thanks to \jwst,\ we can directly measure the metallicity of large samples of very distant sources \citep[e.g.,][]{curti_2023}. In contrast, over the past few years, observations with the Atacama Large Millimeter Array (ALMA) have revealed the existence of dust-rich, and in some cases seemingly over-abundant, galaxies within a few hundred million years of the Big Bang (see \citealt{witstok_2023} for a recent compilation). This has sparked discussions on the creation and maintenance of dust grains in early galaxies. The short timescales involved seem to preclude a dominant role for asymptotic giant branch (AGB) stars, leaving supernovae (SNe) and grain growth in the interstellar medium (ISM) as the most viable alternatives for dust production, contending against destruction via reverse shocks \citep{gall_2011, popping_2017, schneider_2023}. Different flavors of these mechanisms have been incorporated into models and simulations, yet the ever increasing body of observations persistently challenges their predictions.\\

In this work we delve into these subjects, concentrating our investigation on a faint \subl\ galaxy at $z=6.072$, positioned in projection behind the cluster RXJ0600--2007
(\citealt{fujimoto_2021, laporte_2021}, F21 and L21 hereafter; \citealt{sun_2022}). An extensive observational campaign to characterize the full spectrum of this source commenced soon after its discovery \citep{fujimoto_2024} and is still ongoing. Here we present the initial findings from a series of programs conducted with ALMA.\ With the latter we accumulated approximately $\sim$60 hours of observations, which were complemented by \textit{Hubble} Space Telescope (HST) and \jwst\ imaging as well as integral field near-IR spectroscopy from \jwst. Armed with this rich dataset, we also aim to provide a reference for further studies involving larger samples of similar galaxies whose spectroscopic confirmation is now within reach \citep[e.g.,][]{khullar_2021, glazer_2023, fudamoto_2024}.\\ 

Details on the data reduction and analysis are presented in Sections \ref{sec:data} and \ref{sec:analysis}. In Section \ref{sec:results} we report on the obscured SFR, its fraction of the total value, the dust mass and temperature (\mdust\ and \tdust), the cold gas mass and properties as traced by dust, neutral and ionized atomic carbon (C\,{\footnotesize I} and C\,{\footnotesize II}), carbon monoxide (\co), dynamics, and direct estimates of dust-to-stellar and dust-to-gas ratios ($\delta_{\rm DGR}$), and we put them in the context of the literature values for brighter UV galaxies and models at similar redshifts. Our conclusions are presented in Section \ref{sec:conclusions}. The reader can find detailed descriptions of the whole dataset \citep{fujimoto_2024} and a focused analysis of the \jwst\ photometry  \citep{gimenez-arteaga_2024} in companion papers. The study of the properties of the photon-dominated regions traced by multiple far-IR (FIR) lines is left to a dedicated work (Lee et al. in prep.).\\ 

Throughout this work we make use of the AB system to report magnitudes. Unless otherwise specified, we adopt a \cite{chabrier_2003} initial mass function (IMF) and a $\Lambda$ cold dark matter cosmology with $\Omega_{\rm m} = 0.3$, $\Omega_{\Lambda} = 0.7$, and $H_0 = 70\,\mathrm{km\,s^{-1}\,Mpc^{-1}}$.

\section{Data}
\label{sec:data}

\subsection{The target}
\label{sec:target}

The target presented in this study was discovered serendipitously during the mapping of dust emission at
$\lambda_{\rm obs}=1.1\,$mm within and around 33 clusters from legacy surveys using \hst \ and the \textit{Spitzer} Space Telescope
-- the ALMA Lensing Cluster Survey (ALCS; \citealt{kohno_2023,
fujimoto_2021, fujimoto_2023_counts, kokorev_2022}).
Five lensed images of this galaxy are now securely spectroscopically confirmed. In particular, a highly magnified arc composed of two images of a peripheral region of the source (dubbed ``z6.1-6.2''; magnification factor $\mu \sim 100$) crosses a caustic line and stands out as one of the brightest observed \cii$158\,\mu$m emitters at $z>6$ (F21). The emission of a second lensed image (``z6.3''), approximately $15\arcsec$ distant from the arc, is also strongly boosted ($\mu\sim20$) and offers an excellent view of global properties of this galaxy. This has already allowed for investigations of resolved properties in the UV wavelength regime down to $<300$ pc with \hst\ observations. Its total intrinsic absolute UV rest-frame magnitude is $\sim3\times$ fainter than the knee of the luminosity function of Lyman break galaxies at $z=6$ (F21).

\begin{figure*}
    \includegraphics[width=\textwidth]{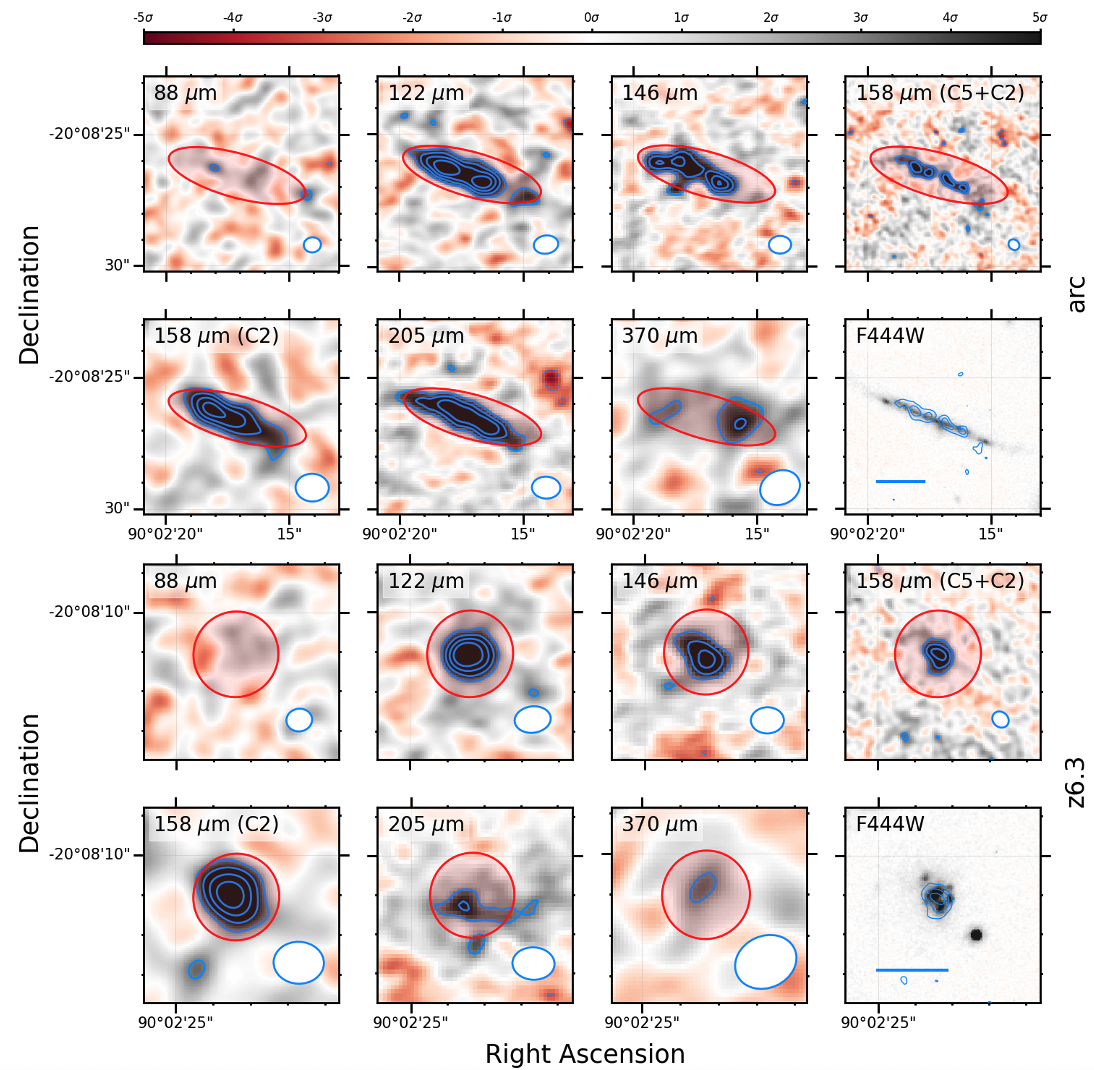}
    \caption{ALMA continuum bands of the arc (top rows, 7.5" side) and the z6.3 image (bottom rows, 5" side). The rest-frame wavelength of each map is labeled. The ALMA images are color-scaled within $\pm5$  times the rms per pixel in each band. Solid (positive) and dashed (negative) blue contours are at $\pm3,\,5,\,7,$ and $9\sigma$.  The solid red line and shaded area indicate the aperture adopted to extract the photometry. The beam size is shown in the bottom-right corner of each cutout. The bottom-right panel of each series shows the \jwst/NIRCam image at 4.5 $\mu$m. In the latter, the point spread function has a full width at half maximum of $0\farcs16$ and the blue contours mark the continuum at $158$ $\mu$m (C5+C2). The blue bar shows a $10$ kpc scale.}
    \label{fig:measurements}
\end{figure*}
\subsection{ALMA multiwavelength observations}
\label{sec:alma_observations}

The goal of our ALMA campaign is to characterize the properties of the cold ISM in this \subl\ galaxy as completely as possible. We thus followed up its lensed images in Bands 3, 5, 6, 7, and 8\footnote{Program IDs:  2021.1.00247.S, 2021.1.00055.S, 2022.1.00195.S (PI: S. Fujimoto), and 2021.1.00181.S (PI: F. Valentino).}, targeting known FIR and submillimeter lines and their underlying dust continuum emission. A summary of observations, their integration time, observed and rest-frame frequency and wavelengths, depths, and beam sizes is provided in Table \ref{tab:obs}.
The ALMA data were consistently reduced and calibrated using the
Common Astronomy Software Applications (CASA) package version
6.2.1.7 (\citealt{mcmullin_2007}) with the standard pipeline script. \textsc{TCLEAN} was used to produce the continuum
maps averaging all channels in the spectral windows not dominated by bright lines (whose expected position is accurately known from the robust redshift estimate). The
\textsc{TCLEAN} routines were executed down to the $2\sigma$ level. We
chose a pixel scale approximately ten times smaller than the beam size and adopted a common spectral channel
bin of $40$ \kms. The final natural-weighted map has
synthesized beams and continuum sensitivities listed in Table \ref{tab:obs}. 
More details about the individual ALMA programs and the data reduction can be found in \cite{fujimoto_2024}.\\

\subsection{Ancillary datasets}
\label{sec:ancillary}
RXJ0600--2007 ($z=0.43$) has been observed as part of the Reionization Lensing Cluster Survey \citep[RELICS;][]{coe_2019} and the Massive Cluster Survey \citep[MACS;][]{ebeling_2001}. It thus benefits from ample UV to near-IR photometric coverage with the HST and \textit{Spitzer}. A homogeneous data reduction of all programs targeting this cluster and a consistent re-extraction of the photometry is presented in \cite{kokorev_2022} as part of the Complete \textit{Hubble} Archive for Galaxy Evolution (CHArGE) initiative. It has been recently complemented by JWST observations with the Near-Infrared Camera (NIRCam; \citealt{rieke_2005}) and Spectrograph (NIRSpec; \citealt{jakobsen_2022}) integral field unit (\#GO 1567, PI: S. Fujimoto). The imaging was taken with the F115W, F150W, F277W, F356W, and F444W filters. The data reduction is performed following the same steps described in \cite{valentino_2023}. NIRCam images of the arc and the z6.3 images at $4.5$ $\mu$m are shown in Figure \ref{fig:measurements}. The integral field spectroscopy came at high spectral resolution with the G395H/F290LP combination ($R\sim2700$, $\lambda_{\rm obs} \sim 2.9-5.1$ $\mu$m), thus covering the redshifted optical rest-frame emission of lines from the ionized gas (notably including H$\alpha$, H$\beta$, \oiii$\lambda\lambda 4959,\,5007$, \nii$\lambda\lambda 6549,\,6584$, \sii$\lambda\lambda 6716,\,6731$, the auroral line \oiii$\lambda 4364$, and many others). For further details about the \jwst\ and \hst\ imaging and spectroscopy, we refer the reader to the companion papers \citep{gimenez-arteaga_2024, fujimoto_2024}.

\section{Analysis}
\label{sec:analysis}

\subsection{Dust continuum photometric extraction}
\label{sec:photometry}
We extracted the photometry within elliptical apertures in the continuum emission ALMA maps. We ran the pythonic version of \textsc{Source Extractor} \citep{bertin_1996}, \textsc{SEP} \citep[v1.2.1,][]{barbary_2016}, to detect the centroid, orientation, and ellipticity of the apertures in each band independently. We measured total flux densities within adaptively scaled \cite{kron_1980} apertures (\textsc{FLUX\textunderscore AUTO} in \textsc{Source Extractor}'s nomenclature). To ensure that we measured flux densities across wavelengths within consistent portions of each lensed image, in every band we fixed our fiducial aperture to that retrieved for Band 7 in the $122$ $\mu$m rest frame, where the S/N of the detection is high ($10$ and $14$ for z6.3 and the arc, respectively). We show the extent of such an aperture in every map in Figure \ref{fig:measurements}. The resulting dust continuum emission flux densities for z6.3 and the arc are reported in Table \ref{tab:measurements}, together with available upper limits on \textit{Herschel} photometry (de-blended and extracted as described in \citealt{sun_2022}). The uncertainties are computed from the rms per beam in each continuum image scaled by the number of independent beams within the aperture \citep[e.g.,][]{bethermin_2020}. 

When observing in Bands 6, 7, and 8, we separately pointed each lensed image. Given the larger field of view in Bands 3 and 5, the pointing was centered between z6.3 and the arc.
In this case, the lensed images are well within the inner portions of the primary beam. Therefore, a correction for the primary beam was not needed. In the case of Band 6, we obtained observations in two configurations (C2 and C5), which allows us to test for possible flux lost on the full arc scale (the maximum recoverable scale for Band 6 in C5 is of $\sim3$"). Independent measurements of the continuum emission in the two configurations are consistent (Table \ref{tab:measurements}). We thus combined these datasets with \textsc{CASA} for our final measurements, improving the signal-to-noise of the detection.

Choosing Bands 5 or 6, where the detections are also robust, as references to determine the aperture instead of Band 7 does not appreciably impact the measurements. Also, we excluded possible strong biases due to widely different spatial distribution of the dust emission across bands by measuring total flux densities independently at each frequency, obtaining consistent results. 

Finally, we counterchecked our aperture photometry against modeling of the emission with the task \textsc{IMFIT} in \textsc{CASA} and single Gaussians profiles. The results are overall in good agreement with those from aperture photometry especially when detections are robust (Table \ref{tab:measurements}). Low signal-to-noise estimates and upper limits are also consistent (e.g., Band 8). We note that the assumption of a single, smooth profile captures reasonably well the global light emission from each lensed image, but clumping on smaller scales is likely present. This is evident from the optical rest-frame emission captured by \jwst\ (Figure \ref{fig:measurements}) and, to a lesser extent, by the highest resolution ALMA map in Band 6. Also, the assumption of a Gaussian light distribution is arguably not apt for the highly stretched arc, where two distinct peaks have been identified in the \cii\ maps at low spatial resolution (F21). Yet, the total flux densities are consistent with those from aperture photometry, which are robust against clumping and spatial inhomogeneities. This suggests that the these spatial variations are not fully captured by ALMA at the current spatial resolution and depths in most bands, but this does not affect our conclusions.

\begin{table*}[]
    \centering
    \caption{ALMA continuum photometry and Band 3 line measurements.}
    \begin{tabular}{lccccc}
    \toprule
    \toprule
    & $\lambda_{\rm rest}$ & \multicolumn{4}{c}{$\mu F_{\nu}$} \\
    \textit{Herschel}  & \small $[\mu\mathrm{m}]$ & \multicolumn{4}{c}{\small [mJy]}  \\ 
    \midrule
    SPIRE 250 $\mu$m & 35 & \multicolumn{4}{c}{$<15.1$} \\ 
    SPIRE 350 $\mu$m & 50 & \multicolumn{4}{c}{$<16.1$} \\ 
    SPIRE 500 $\mu$m & 70 & \multicolumn{4}{c}{$<18.0$} \\ 
    \midrule
    & & \multicolumn{2}{c}{z6.3}& \multicolumn{2}{c}{arc}\\ \cmidrule{3-4} \cmidrule{5-6}
    ALMA &  & Aperture & Model  & Aperture & Model  \\
    \midrule
    Band 8 & 88  & $0.12 \pm 0.89$ & $0.96 \pm 0.52$ & $0.66 \pm 1.23$ & $1.99 \pm 0.62$ \\
    Band 7 & 122 & $0.50 \pm 0.08$ & $0.36 \pm 0.05$ & $0.82 \pm 0.11$ & $0.84 \pm 0.09$\\
    Band 7 & 146  & $0.38 \pm 0.09$ & $0.32 \pm 0.07$ & $0.76 \pm 0.12$ & $0.71 \pm 0.10$  \\
    Band 6 & 158 & $0.25 \pm 0.08$ & $0.20 \pm 0.03$ & $0.49 \pm 0.11$ & $0.52 \pm 0.07$\\
    Band 5 & 205 & $0.12 \pm 0.03$ & $0.13 \pm 0.03$ & $0.24 \pm 0.04$ & $0.23 \pm 0.02$ \\
    Band 3 & 370 & $0.008 \pm 0.007$ & $0.008 \pm 0.003$ & $0.04 \pm 0.01$ & $0.06 \pm 0.01$ \\
    \midrule
    \multirow{3}{*}{Band 3 (CO, \ci)} & $\mu I$ \small[\jykms]& \multicolumn{2}{c}{$<0.013$}&  \multicolumn{2}{c}{$<0.046$}  \\   
     & $\mu L'$ \small [\kkmspc] & \multicolumn{2}{c}{$<3.2\times10^{8}$} & \multicolumn{2}{c}{$<1.1\times10^{9}$} \\  
     & $\mu L$ \small [\lsun] & \multicolumn{2}{c}{$<5.5\times10^{6}$} &  \multicolumn{2}{c}{$<1.9\times10^{7}$} \\  
    \bottomrule
  \end{tabular}
  \tablefoot{\smallskip All estimates are not corrected for lensing. All upper limits are at $3\sigma$.  
  \textit{Herschel}/SPIRE from \cite{sun_2022}. In Band 6, for z6.3: $\mu F_{\nu, \rm aper} = (0.25\pm0.06)$ and $(0.18\pm0.12)$ mJy for C2 and C5, respectively. For the arc: $\mu F_{\nu, \rm aper} = (0.37\pm0.08)$ (C2) and $(0.52\pm0.16)$ mJy (C5). Band 3 lines: \citwo\ ($\lambda_{\rm rest}=370.41\,\mu$m, $\nu_{\rm rest}=809.34\,$GHz), \coseven\ ($\lambda_{\rm rest}=371.65\,\mu$m, $\nu_{\rm rest}=806.65\,$GHz). Upper limits over $\Delta v=180\,$\kms\ and estimated from spectra binned at $dv=40\,$\kms.}
    \label{tab:measurements}
\end{table*}

\subsection{Spectral extraction}
\label{sec:spectra}
\begin{figure}
    \centering
    \includegraphics[trim={0cm 1cm 3cm 3cm}, width=\columnwidth]{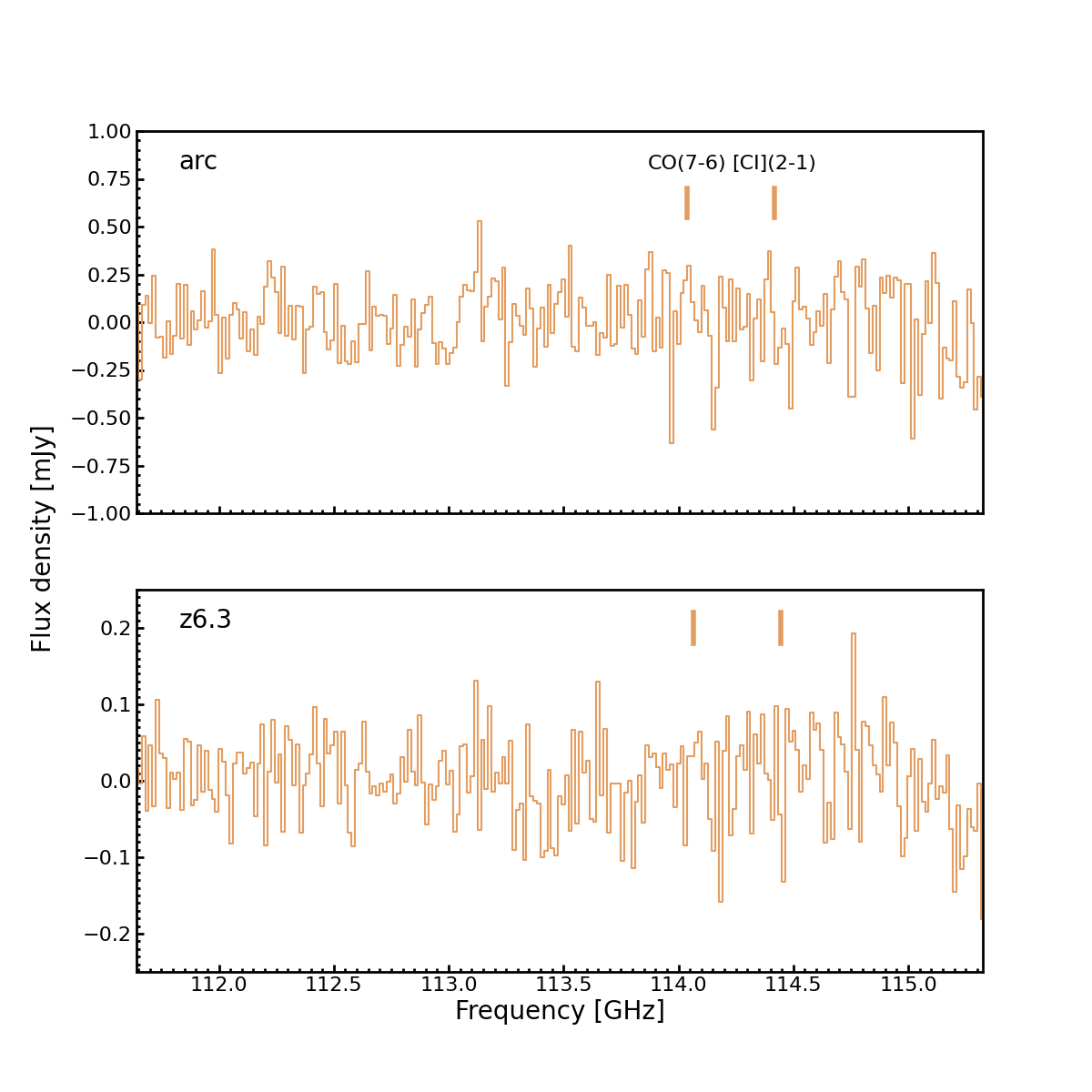}
    \caption{Observed Band 3 spectra of the arc (top) and the z6.3 image (bottom). The expected positions of the \citwo\ and \coseven\ lines are marked.}
    \label{fig:spectrum}
\end{figure}
We extracted the spectra of the lensed images over the same apertures used to measure the continuum emission in order to derive consistent ratios. Here we focus particularly on Band 3 observations targeting the \citwo\ ($\nu_{\rm rest}=809.34\,$GHz) and \coseven\ ($\nu_{\rm rest}=806.65\,$GHz) lines, proxies of the molecular gas of the galaxy, and we defer detailed analyses of photon-dominated regions to future work (Lee et al. in prep.). We show the extracted Band 3 spectra around the expected location of the \ci\ and CO line emission in Figure \ref{fig:spectrum}. We do not detect any line emissions. We thus set upper limits on the fluxes integrated on the \cii\ line width from low spatial resolution measurements ($\Delta v=\mathrm{FWHM}=180$ \kms, F21). Based on spectra binned at $dv=40$ \kms, we obtain $3\sigma$ upper limits on the line fluxes as $I_{\rm line}=3\times \mathrm{rms} \sqrt{\Delta v\,dv}$, where rms is the aperture noise per channel in the spectral window where the lines fall. We applied a mild clipping at $3\sigma$ to remove outliers, which resulted in a $6$\% lower rms. The limits on line fluxes and luminosities are reported in Table \ref{tab:measurements}. We note that the upper limits on z6.3 and the arc differ because of the areas that we considered to estimate the rms.    

\begin{table}
    \small
    \centering
    \caption{Physical properties.}
    \begin{tabular}{lcc}
    \toprule
    \toprule
    & z6.3 & arc\\ 
    \midrule
    R.A. \small [h:m:s]& 06:00:09.55 &  06:00:09.13\\
    Decl. \small [d:m:s]& $-$20:08:11.26 &      $-$20:08:26.49\\
    $z_{\rm spec}$ & $6.0719$ & $6.0736$\\
    $\mu$ & $21^{+14}_{-7}$ &   $163^{+27}_{-13}$${\dagger}$\\ 
    log($\mu M_\star$) \small [$M_\odot$] & $10.2\pm0.2$ & $10.6\pm0.2$ \\
    $\mu L_{\rm FIR, 42.5-122.5\,\mu\mathrm{m}}$ \small [$L_\odot$]& 
 $7.1^{+6.1}_{-3.5} \times 10^{11}$ & $5.7^{+4.1}_{-2.1} \times 10^{11}$ \\
    $\mu L_{\rm IR, 8-1000\,\mu\mathrm{m}}$ \small [$\mu L_\odot$]& 
 $1.1^{+2.2}_{-0.6} \times 10^{12}$ & $0.7^{+0.6}_{-0.2} \times 10^{12}$ \\
    $\mu L_{\rm IR,\,CO(7-6)}$ \small [$L_\odot$]& 
 $<0.4\times 10^{12}$ & $<1.3\times 10^{12}$ \\
    $\mu M_{\rm dust}$ \small [$M_\odot$] & $0.4^{+0.4}_{-0.2} \times 10^{8}$ & $1.6^{+1.7}_{-0.8} \times 10^{8
}$ \\
    \tdust\ \small [K] & $50^{+20}_{-13}$ & $37^{+11}_{-7}$\\
    $\beta_{\rm IR}$ & $1.8^{+0.2}_{-0.2}$ & $1.6^{+0.2}_{-0.2}$\\
    $\mu$SFR(IR) \small [\myr]& $164^{+334}_{-95}$ & $111^{+96}_{-37}$ \\
    $\mu$SFR(IR, \coseven) \small [\myr]& $<53$ & $<193$ \\
    $\mu$SFR(FUV) \small [\myr]& $49^{+7}_{-6}$& --  \\
    Total $\mu$SFR(FUV$_{\rm corr}$) \small [\myr]& $135^{+171}_{-48}$  & -- \\
    Total $\mu$SFR(\ha$_{\rm corr}$)/$k_Z$ \small [\myr]& $141^{+42}_{-14}$  & -- \\
    $\mu M_{\rm gas}$(\ci) \small [\msun]& $<0.1-1\times10^{10}$ & -- \\
    $\mu M_{\rm gas}$(\cii)$_{\rm Z18}$ \small [\msun]& $(6.9\pm0.6)\times10^{10} $ & -- \\
    $\mu M_{\rm gas}$(\cii)$_{\rm H21}$ \small [\msun]${\mathsection}$& $2.2^{+1.5}_{-0.9}\times10^{11}$ & -- \\
    $M_{\rm dyn}$ \small [\msun]$\ddag$& $3^{+1}_{-1} \times10^{9}$ & -- \\
    \bottomrule
  \end{tabular}
  \tablefoot{\smallskip All estimates are not corrected for lensing (excluding the dynamical mass). The estimates of physical properties from optical and FIR/submillimeter SED modeling are expressed as the median value and the 16-84\% of their posterior distributions. Upper limits at $3\sigma$. 
  Total $\mu$SFR(FUV$_{\rm corr}$), $\mu$SFR(\ha$_{\rm corr}$) are derived from the luminosities at 1500\AA\ and of the \ha\ line corrected for the dust attenuation as in \cite{hao_2011} and \cite{murphy_2011} (Section \ref{sec:unobscured_sfr}). $\mu$SFR(\ha$_{\rm corr}$) is decreased by a factor of $k_Z=2.5$ with respect to the calibration in \cite{murphy_2011} to account for the low metallicity ($12+\mathrm{log(O/H)}=(8.11\pm0.20)$ or $Z\sim0.25$ $Z_\odot$, \citealt{kennicutt-evans_2012, theios_2019, shapley_2023}).\\
  ${\dagger}$ For the arc, this factor is defined as the observed
luminosity of the image divided by the local luminosity of the strongly lensed sub region near the caustic line shown in Figure 4 of F21. The global $\mu$ factor for the whole galaxy (i.e., the observed luminosity of the multiple image divided by the overall intrinsic luminosity of the galaxy) is $\mu_{\rm whole}=29^{+4}_{-7}$ (F21).\\
${\mathsection}$ The calibration in \cite{heintz_2021} returns the \hi\ gas mass.\\
$\ddag$ From \cii\ observations in F21.}
    \label{tab:best-fit}
\end{table}

\subsection{Long-wavelength spectral energy distribution modeling}
\label{sec:fir_sed}
\begin{figure*}
    \includegraphics[width=\textwidth]{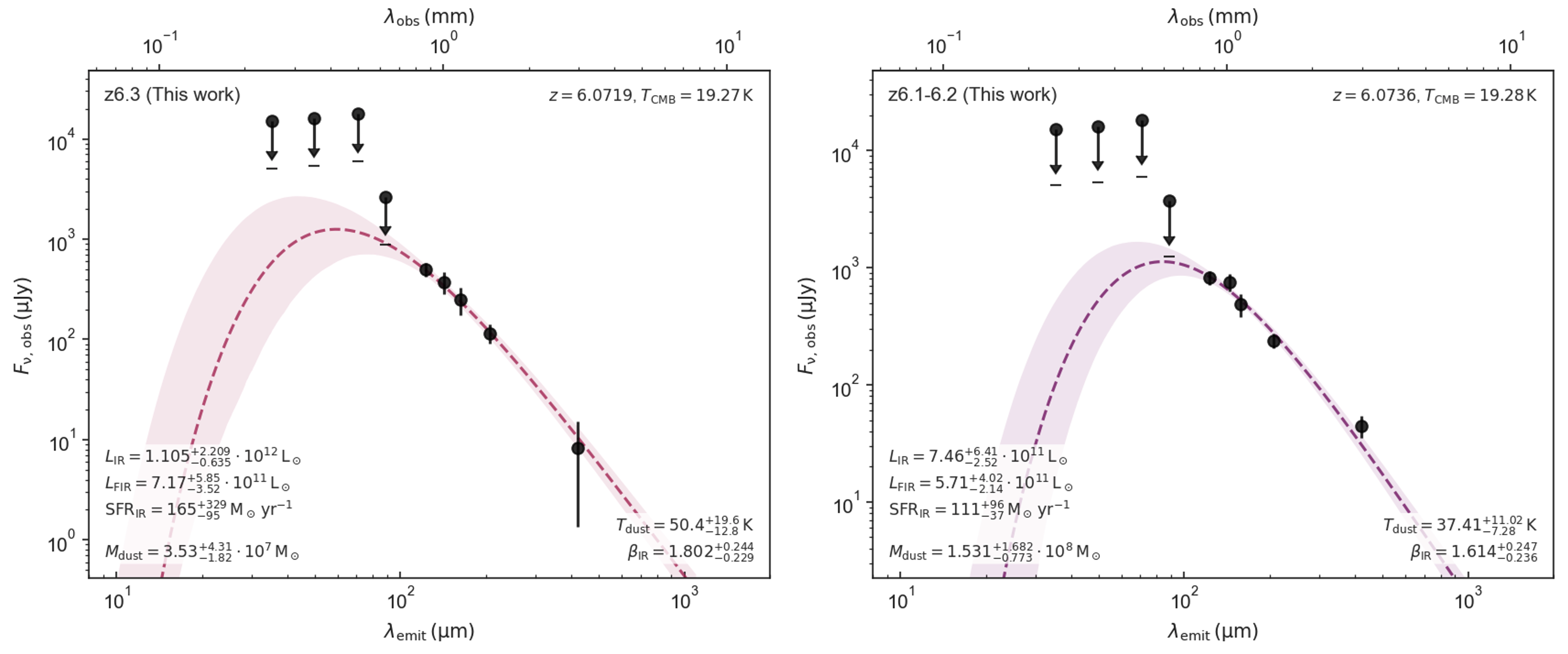}
    \caption{FIR SED modeling for z6.3 (\textit{left}) and z6.1-6.2 (arc, \textit{right}). Black circles and arrows indicate detections and $3\sigma$ upper limits on the observed photometry. The dashed line and shaded areas mark the best-fit model with \textsc{Mercurius} under the assumption of an optically thin emission.}
    \label{fig:sed_free}
\end{figure*}

We modeled the ALMA and \textit{Herschel} photometry using the Bayesian code \textsc{Mercurius} \citep{witstok_2022, witstok_2023}. We assumed optically thin dust emission and initially left the dust temperature (\tdust) and emissivity slope ($\beta$) free to vary. For consistency with the literature compilation presented in \cite{witstok_2023}, a primary reference later in this work, we adopted their same priors. For the dust temperature, we used the default gamma distribution with shape parameter $a=1.5$ and shifted it to start at the temperature of the cosmic microwave background (CMB, $T_{\rm CMB}$); for $\beta$, we imposed a Gaussian prior centered at $1.8$ and with a standard deviation of $0.25$. These priors reflect the belief that \tdust\ is less likely to be extremely high and the known distribution of $\beta$ across redshifts \citep{witstok_2022, witstok_2023}. The effect of the CMB is taken into account in the modeling \citep{dacunha_2013}. We adopt the built-in dust emissivity coefficient $\kappa = \kappa_0 (\nu/\nu_0)^{\beta}$ with $\kappa_0 = 8.94\,\mathrm{cm^{2}\,g^{-1}}$ at $\nu_0=1900\,\mathrm{GHz}$ ($\sim 158\,\mu$m, \citealt{hirashita_2014}), consistent with the choice in several recent works \citep{schouws_2022, witstok_2022, witstok_2023, valentino_2022}. The choice of different emissivity coefficients has an obvious systematic impact on the final fit estimates with differences amounting up to an order of magnitude. The best-fit models for z6.3 and the arc are shown in Figure \ref{fig:sed_free} and the parameters are reported in Table \ref{tab:best-fit}. We also attempted alternative models at fixed $\beta=1.5, 2$ -- typical choices when fewer data points are available -- and implementing a self-consistent calculation of the wavelength $\lambda_0$ at which the dust emission becomes optically thin (i.e., optical depth $\tau=1$), given the knowledge of the intrinsic source size from high-spatial resolution measurements (effective radius $R_{\rm eff}=1.2^{+1.4}_{-0.1}\,$kpc measured in the source plane after de-lensing, F21). The results are overall consistent with the free parameter fit. On the one hand, the absence of strong observational constraints at short wavelengths does not allow for a robust determination of \tdust, as expected. On the other hand, the multiple ALMA detections in the Rayleigh-Jeans tail of the dust emission better constrain \mdust. The range spanned by the best-fit solutions, also subject to the choice of the priors, reflects the systematic uncertainties hindering the modeling of the long-wavelength spectral energy distributions (SEDs) -- here and in the rest of the literature. For completeness and transparency, we report on a different selection of priors and treatment of the optical depth in Appendix \ref{app:alternative_photometries}.\\

We estimate \tdust$\sim 37$ and $\sim50$ K for the arc and z6.3, respectively. Both are consistent with typical \tdust\ values at $z\sim6$ for more massive UV-, optical-, and IR-selected sources \citep{witstok_2023}. We note that, in principle, gradients in \tdust\ or dust composition (thus, $\beta$) could be present \citep{akins_2022} and potentially detectable, given the lensing configuration. As described in F21, the arc strongly magnifies ($\mu_{\rm local}=163^{+27}_{-13}$) a peripheral region of the source plane and it can be used to gauge possible gradients within the galaxy. On the contrary, z6.3 is the brightest lensed image that captures the global properties of the intrinsic source. Marginally higher \tdust\ are indeed preferred for z6.3, suggesting the presence of a negative dust temperature gradient.
However, in practice the resolution and S/N of the integrated detections in crucial ALMA bands at the shortest and longest wavelengths do not allow for a full-fledged resolved analysis at this stage and current speculations on the presence of gradient will have to be tested against new observations. For reference, in the rest of this work we make use of the best-fit parameters with free $\beta$ under the optically thin assumption and we focus on the global properties of the galaxy from the z6.3 image. For completeness, the measurements for the arc are shown in figures and reported in tables.

\subsection{Modeling of optical and near-IR observations}
We modeled the \jwst\ photometry as described in detail in the companion paper by \cite{gimenez-arteaga_2024} and previous works \citep{gimenez-arteaga_2023}. Briefly, we first point-spread-function-matched the observations to the lowest available resolution in NIRCam F444W ($\mathrm{FWHM=0\farcs16}$) and then we modeled the emission with \textsc{Bagpipes} \citep{carnall_2018}. Since we cannot consistently resolve the molecular gas and dust emission within the galaxy with ALMA (especially that at the longest wavelengths to determine \mdust), we made use of the SED modeling of the integrated fluxes. However, we encourage the reader interested in the details of a pixel-by-pixel analysis of z6.3 to consult the dedicated companion paper. \\

We adopted \cite{bruzual_2003} single stellar population models\footnote{As detailed in \cite{gimenez-arteaga_2024}, a \cite{kroupa_2001} IMF was used to model the \jwst/NIRCam photometry. The stellar mass estimates are similar to those derived using a \cite{chabrier_2003} prescription.}, a grid of \textsc{cloudy} \citep{ferland_2017} emission line grids extended to higher ionization parameters (up to log$(U)=-1$), and a \cite{calzetti_2000} attenuation curve. We imposed uniform priors on the attenuation ($A_{V}=0-3$), metallicity ($Z=0-Z_\odot$), formed stellar mass ($M=10^5-10^{11}\,M_\odot$). We finally adopted the stellar mass estimate obtained with a double power-law star formation history (SFH) that is a robust choice against the ``outshining'' effect of young stellar populations (Section \ref{sec:gas_masses}). The \mstar\ estimate is consistent with those from other parameterizations of the SFHs on resolved scales, while that of SFR is notoriously dependent on the shape of the formation history and the time window to average for the calculation. We thus refrained from using the SFR from optical/near-IR SED modeling in this work. 

Finally, we derived the gas-phase metallicity and unobscured SFR on global scales by integrating the measurements of the ionized line emissions in the NIRSpec spectrum \citep{fujimoto_2024} and the stellar continuum in the far-UV (FUV) range of the z6.3 image. The former, expressed as the oxygen gas-phase abundance [12+log(O/H)], was derived via the ``direct method'' based on the electron temperature $T_{\rm e}$ and the detection of the [O\,{\footnotesize III}]$\lambda$4364 auroral line. The global resulting metallicity is $\mathrm{12+log(O/H)}=8.11 \pm 0.20$ ($Z\sim0.25 \,Z_\odot$ for a solar abundance of $\mathrm{12+log(O/H)}=8.69$; \citealt{asplund_2009}). In Section \ref{sec:dust_obscured_sfr}, we return to a discussion of the SFR from the \ha\ Balmer line, sensitive to the formation of stars on timescales of $\lesssim10$ Myr.

\section{A view of the cold gas and dust properties of a \subl\ galaxy at $z=6$ }
\label{sec:results}

\begin{figure}
    \centering
    \includegraphics[width=\columnwidth, trim={0 3cm 0 4cm}, clip]{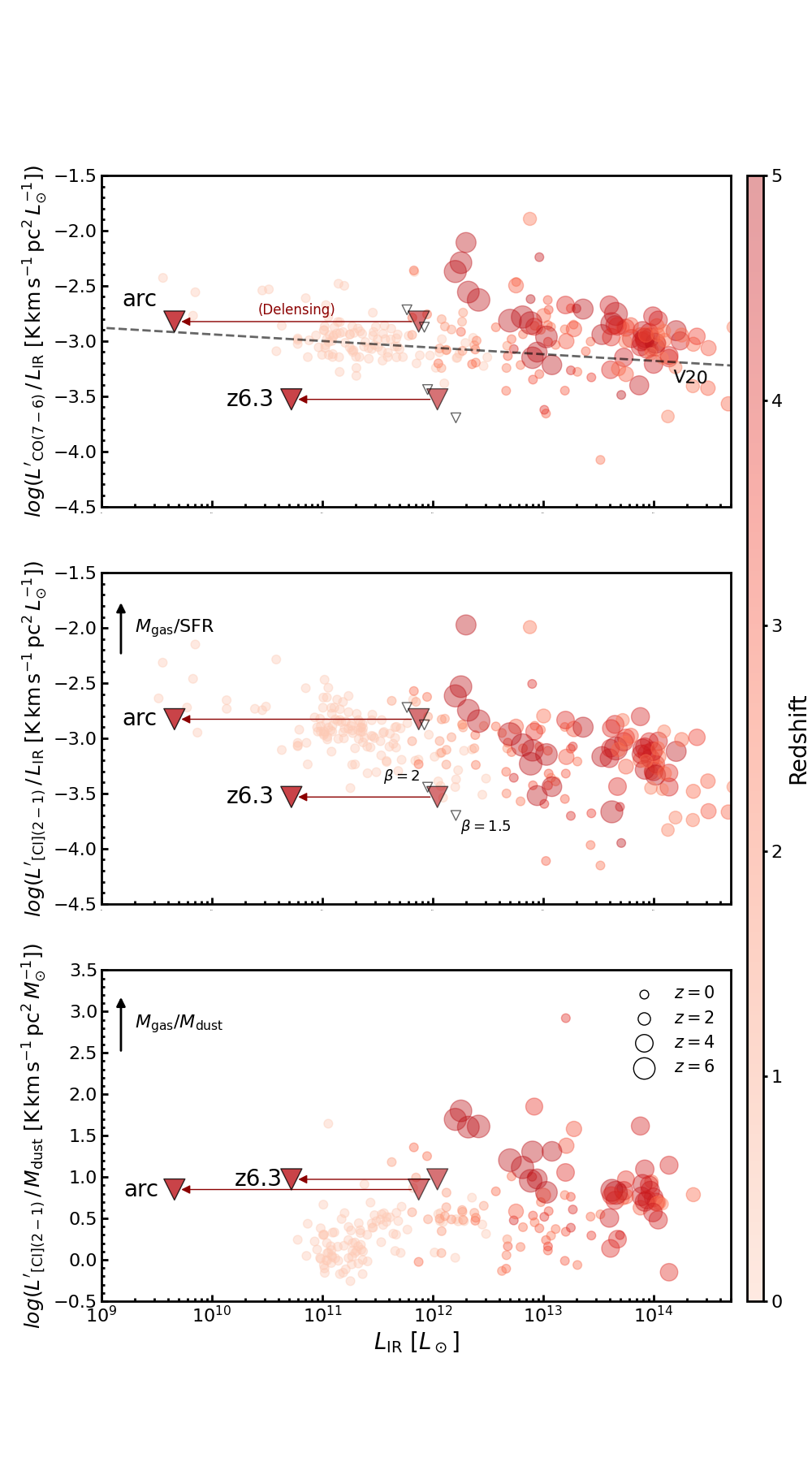}
    \caption{Observed $L'/L_{\rm IR}$ for \coseven\ (\textit{top}), \citwo\ (\textit{center}), and \lprimecitwo/\mdust\ ratios (\textit{bottom}) as a function of \lir\ for our targets, as labeled (downward triangles; $3\sigma$ upper limits) and a literature compilation described in Appendix \ref{app:literature} (red circles; colors and sizes scale with redshift). The  red arrows connect the observed and de-lensed estimates of \lir\ for our targets.
    The open triangles show the upper limits on our targets when fixing $\beta=1.5-2$ in the SED modeling. The dashed line in the top panel indicates the empirical \lir-\lprimecoseven\ relation from \cite{valentino_2020c}. The black arrows in the central and bottom panels indicate the direction of increasing depletion timescales (\lprimecitwo/\lir $\propto$ \mgas/SFR) and the dust-to-gas mass ratio (\lprimecitwo/\mdust $\propto$ $\delta_{\rm DGR}^{-1}$), respectively.}
    \label{fig:lir_lprime-lir}
\end{figure}

\subsection{Gas and dust emission tracers}
\label{sec:observables}
We show ALMA-derived observables (or closely related quantities) in context of what is available in the literature in Figure \ref{fig:lir_lprime-lir}. The data compilation is briefly described in Appendix \ref{app:literature}. In the space of observables, the upper limits on \coseven\ and \citwo\ are consistent with the loci of $L'$/\lir\ ratios (and their scatter) of variously selected IR-emitting galaxies spread across the last 13 Gyr of cosmic time. This is true also for the \lprimecitwo/\mdust\ ratio, both tracers of the cold gas mass in galaxies. The combination of lensing and deep observations allows for the exploration of \lir\ and line ratios typical of local IR-detected objects -- but at the end of reionization. This is a jump of $\sim2$ orders of magnitude in intrinsic \lir\ at $z\sim6$. The exact location of the intrinsic values in Figure \ref{fig:lir_lprime-lir} is of course dependent on the lensing models and it is affected by their uncertainties (Table \ref{tab:best-fit}), but these are confidently smaller than the jump in \lir. The ratios of line and continuum emissions are also affected by uncertainties on differential lensing, but this effect seems limited especially for z6.3 \citep{fujimoto_2024}, our anchor to derive the global properties of this galaxy. 

The ratios shown in Figure \ref{fig:lir_lprime-lir} start telling us the story of our target. In broad terms, \coseven\ traces the dense and warm star-forming molecular gas and linearly correlates with \lir, a classical tracer of the obscured SFR. The \citwo\ luminosity correlates with the total molecular gas content \citep{yang_2017,valentino_2020b}, despite its exact calibration being sensitive to (here unknown) excitation corrections \citep{dunne_2022}. The \lprimecitwo/\lir\ ratio is thus a proxy for the depletion timescale in the galaxy ($\tau_{\rm depl}=M_{\rm gas}/\mathrm{SFR}$). Since the dust mass is also a tracer of the cold gas mass in galaxies \citep{magdis_2012}, the \lprimecitwo/\mdust\ ratio can serve as a proxy of the (inverse of the) dust-to-gas ratio ($\delta_{\rm DGR}$). Figure \ref{fig:lir_lprime-lir} thus suggests that  the $z=6$ \subl\ galaxy is generally consistent with the observed or extrapolated properties and trends of IR-detected galaxies in terms of depletion timescales and dust-to-gas ratios as a function of the obscured SFR.

\subsection{Gas and dust masses}
\label{sec:gas_masses}
We computed the total gas mass of our galaxy by leveraging the measurements available for z6.3. We converted the \citwo\ upper limit to \mgas\ as \citep{papadopoulos-greve_2004}\begin{equation}
M_{\rm gas}(\mathrm{[CI]}) = 1375.8 \times 10^{-12} \, 
            \frac{D_{\rm L}^2 \, Q_{21}\,S_{\rm \nu}\,\Delta v}{(1+z)\,A_{21} X_{\rm [CI]}}
,\end{equation}
where $D_{\rm L}$ is the luminosity distance in Mpc, $Q_{21}$ is the excitation factor, $S_{\rm \nu}\,\Delta v$ the upper limit on the velocity-integrated line flux, $A_{21}=2.68\times10^{-7}$ s$^{-1}$ the Einstein coefficient, and $X_{\rm [CI]}$ the neutral atomic carbon abundance. The excitation correction is a major factor of uncertainty and it is unknown for this object. In order to obtain an upper limit on \mgas, we considered $Q_{21}=[0.05-0.4]$, the range expected for $T_{\rm dust}\sim50$ K \citep{papadopoulos_2022}. Similarly, we do not know the abundance $X_{\rm [CI]}$. We thus adopted the value for lower redshift main-sequence objects with solar-like metallicity ($\sim 1.5 \times 10^{-5}$, \citealt{valentino_2018}, similar to the average obtained by \citealt{dunne_2022} for their literature sample) and assumed that it scales linearly with metallicity to our $\sim25$\% $Z_{\odot}$ estimate from NIRSpec \citep{glover_2016, heintz_2020}. For reference, the metallicity scaling in \cite{heintz_2020} has an intrinsic scatter of $\sim0.2$ dex. With these variety of assumptions, the upper limit on \lprimecitwo\ is converted into $\mu M_{\rm gas} < (0.1 - 1.0)\times10^{10}$ \msun.\\

We calculated an alternative estimate of the gas mass from the \cii\ emission. This bright line is well detected in our source (F21, Lee et al. in prep.), so the uncertainty must be due to the calibration itself. By assuming the conversion factor in \cite{zanella_2018}, we find $\mu M_{\rm gas}(\mathrm{[CII]})=30\times L_{\rm CII} = (6.9\pm0.6)\times10^{10}$ \msun\ (F21). More recently, lower conversion factors for specific galaxy populations have also been suggested \citep[$\alpha_{\rm [CII]} \lesssim 10$ \msun/\lsun][]{rizzo_2021, sommovigo_2021}. \cite{heintz_2021} proposed a metallicity-dependent calibration to the neutral atomic gas mass $M{\rm(HI)}$ based on a sample of local dwarf galaxies and distant gamma-ray bursts with \cii\ in absorption in the rest-frame UV. By applying this conversion, we obtain $\mu M_\mathrm{HI} (\mathrm{[CII]}) = 2.2^{+1.5}_{-0.9}\times10^{11}$ \msun, where the uncertainties include those on the metallicity and the calibration.\\

The gas dynamics derived in F21 offer an upper limit on \mgas\ independent of the assumptions on these calibrations. The subtraction of the intrinsic \mstar\ from the dynamical mass estimate $M_{\rm dyn} = (3\pm1)\times10^9$ \msun\ leaves $(2\pm1)\times10^9$ \msun\ for all the remaining components. Neglecting dark matter and dust as major contributors to the total mass of the central disk, this value can be ascribed to cold gas. The upper limits from \citwo\ are consistent with the dynamical estimate ($M_{\rm gas}(\mathrm{[CI]}) < (0.6 - 4.8)\times10^{8}$ \msun) and so is $M_{\rm gas}(\mathrm{[CII]}) = (3\pm1)\times10^9$ \msun\ following \cite{zanella_2018} (F21). The metallicity-dependent calibration from \cite{heintz_2021} exceeds the current limit set by dynamical arguments ($M_{\rm HI}(\mathrm{[CII]}) = 1.0^{+0.7}_{-0.4}\times10^{10}$ \msun). However, it is useful to remind that the distribution and extension of \cii\, \hi, and H$_{2}$ gas might differ. An extended \hi\ disk might not be accounted for by the dynamical modeling as traced by the \cii\ emission. Also, the calibration of \hi(\cii) partially relies on absorption measurements in gamma ray bursts. These measurements are sensitive to the integrated column density and might be representative of the dense central cores of galaxies. These factors should be borne in mind when comparing these calibrations (see \citealt{heintz_2023} for a discussion in the case of a similar low-metallicity galaxy at high redshift).\\

\begin{figure}
    \centering
    \includegraphics[width=\columnwidth]{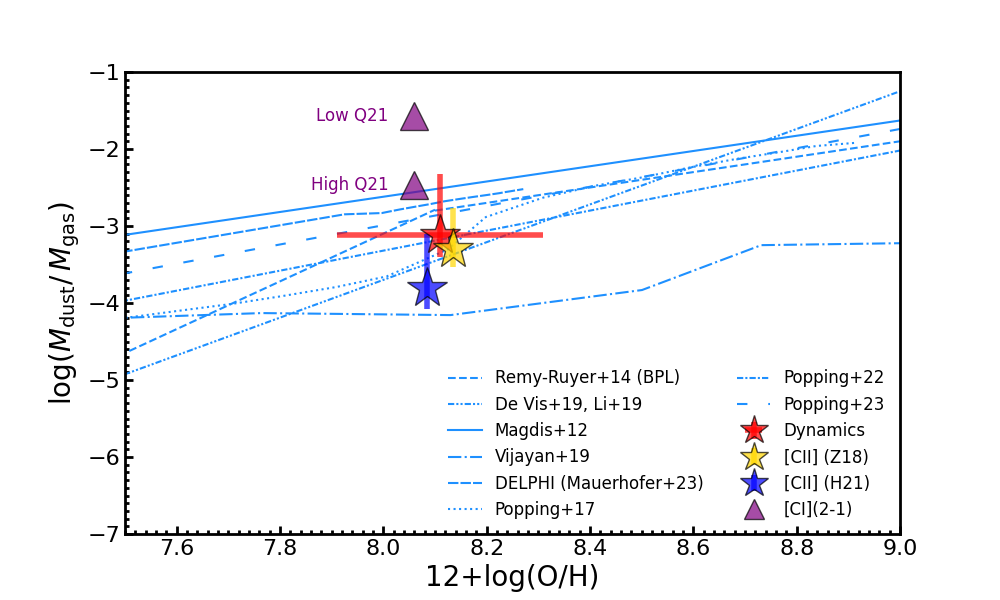}
    \caption{Dust-to-gas mass ratio as a function of metallicity. Purple triangles indicate the $3\sigma$ lower limits from \citwo\ under high- and low-excitation conditions. The gold and blue stars indicate \mgas\ from \cii\ using the conversions in \cite{zanella_2018} and \cite{heintz_2021}, respectively. The red star shows the maximum amount of gas allowed by \cii\ dynamics (F21). The points are slightly offset along the X axis for clarity. The uncertainty on the metallicity is represented by the horizontal red bar.
    The blue lines mark the (extrapolated) relations from \cite{magdis_2012}, \cite{remy-ruyer_2014}, \cite{de-vis_2019}, \cite{popping_2022}, and \cite{popping_2023}, and the models from \cite{mauerhofer_2023}, \cite{popping_2017}, \cite{li_2019}, and \cite{vijayan_2019}.}
    \label{fig:dgr}
\end{figure}

Figure \ref{fig:dgr} shows the global dust-to-gas mass ratio, $\delta_{\rm DGR}$, as a function of metallicity based on the \mgas\ estimates described above. The upper limits on $M_{\rm gas}\mathrm{([CI])}$ (under two sets of excitation conditions) reflect the location of z6.3 in the \lir-\lprimecitwo/\mdust\ ratio shown in Figure \ref{fig:lir_lprime-lir}. For the fiducial galaxy-integrated metallicity $\mathrm{12+log(O/H)=(8.11\pm0.20)}$ measured via the direct method, the $\delta_{\rm DGR}$ is generally consistent with models and empirical relations in the literature (\citealt{remy-ruyer_2014, de-vis_2019,magdis_2012, vijayan_2019, mauerhofer_2023, popping_2017, popping_2022, popping_2023}; furthermore, the results from \citealt{de-vis_2019} are consistent with the predictions from \citealt{li_2019}). We note that no strong evolution in the $\delta_{\rm GDR}$-metallicity relation is seen as a function of redshift and estimates from distant absorption systems are in agreement with local measurements \citep{peroux_2020, popping_2022, heintz_2023_dtg}. Reversing the argument, $\delta_{\rm DGR}\sim0.001$ ($\sim10\times$ lower than typical values of massive dusty star-forming galaxies, \citealt{magdis_2012}) is consistent with the detection of \cii\ and upper limits on \citwo, under the assumptions on the calibrations mentioned above.\\ 

However, we note that the uncertainties on the measurements, the intrinsic scatter of the relations involved in Figure \ref{fig:dgr}, and the fact that we are reporting measurements for a single object certainly do not allow us to exclude any models or trends. 
Also, given the original blind detection in \cii\ and dust, we might be selecting an intrinsically dust-rich object -- yet, consistent with the scaling relations shown so far.
In addition, as for every global measure in this and common in the literature, we are averaging measurements on large spatial regions. The derived quantities are thus luminosity-weighted and this might introduce biases \citep[an ``outshining'' effect,][]{sawicki_1998, papovich_2001, gimenez-arteaga_2023}. In this case, the most metal-poor and least dust-obscured regions dominate the emission of the \oiii\ auroral line used to estimate the metallicity. These might not be the same regions emitting in the FIR. There are indeed hints of different FIR and optical emission distributions when we compare the highest ALMA resolution maps and those from \jwst\ (Figure \ref{fig:measurements}) and across the \jwst\ photometry itself \citep{gimenez-arteaga_2024}. The geometric distributions of gas (H$_{2}$, \hi, and \hii) and dust and their spatial overlap (or lack-of) are factors in the calculation of $\delta_{\rm DGR}$. However, we are not in the position to fully explore the presence of gradients in the ALMA maps, given their resolution. Future work on higher spatial resolution data will trace the dust distribution within the galaxy and allow for the quantification of the ``outshining'' effect at long wavelengths. 

\subsection{Dust-obscured SFR}
\label{sec:dust_obscured_sfr}
The SED modeling described in Section \ref{sec:fir_sed} returns \lir, a direct tracer of the dust-obscured SFR(IR). The latter is computed as $\mathrm{log(SFR(IR)} / M_\odot\, \mathrm{yr}^{-1}) = \mathrm{log}(L_{\rm IR} / \mathrm{erg\, s^{-1}}) - 43.41$ \citep{murphy_2011, kennicutt-evans_2012}. 
We did not apply any further correction to homogenize the \cite{kroupa_2001} and \cite{chabrier_2003} IMFs, given their similarity. The significant uncertainties on \tdust\ propagate on \lir\ and on SFR(IR) (Table \ref{tab:best-fit}). Given the well-established linear correlation between \lir\ and \lprimecoseven \citep{greve_2014, liu_2015, lu_2017, kamenetzky_2016}, we derived an alternative constraint on SFR(IR) from the upper limit on \coseven\ (Figure \ref{fig:lir_lprime-lir}). For this calculation, we adopted the parameterization:
\begin{equation}
\begin{split}
\mathrm{log}(L'_{\rm CO(7-6)} / [\mathrm{K\,km^{-1}\,s^{-1}\,pc^2}]) = \, & (0.94 \pm 0.01) \times \mathrm{log}(L_{\rm IR} / L_\odot) + \\
\quad \, &  (-2.34\pm 0.09)
\end{split}
\end{equation}
with an intrinsic scatter of 0.16 dex  \citep{valentino_2020c}. The $3\sigma$ upper limit on $\mu$SFR(IR, CO(7-6))$<53$ \myr\ is $\sim3\times$ more stringent than the loosely constrained central value from the SED modeling for z6.3 ($\mathrm{\mu SFR_{IR}} = 164^{+334}_{-95}$ \myr), but consistent with it. The higher upper limit derived for the arc, which is a direct consequence of the larger projected area for the flux integration (Section \ref{sec:spectra}), is also consistent with the estimate from the FIR photometry. We note that the lower estimates of \lir\ and SFR(IR) derived by rescaling Band 6 observations in F21, L21, and \cite{sun_2022} are in full agreement with those in Table \ref{tab:best-fit} -- unsurprisingly, at least for the arc image whose best-fit \tdust$\sim 37$ K and $\beta\sim1.6$ are similar to the shape assumed in these works.

\subsection{Unobscured and total SFR}
\label{sec:unobscured_sfr}
We computed the unobscured SFR from the FUV stellar continuum of z6.3 captured by \jwst. We estimated the $1500$\AA\ rest-frame emission by convolving our blueshifted best-fit modeling of NIRCam imaging with the GALEX FUV filter and converting it into $\mathrm{\mu SFR(FUV)} = 49^{+7}_{-6}$ \myr as $\mathrm{log(SFR(FUV)} / M_\odot\, \mathrm{yr}^{-1}) = \mathrm{log}(L_{\rm FUV} / \mathrm{erg\, s^{-1}}) - 43.35$ \citep{murphy_2011}. We obtain a similar $\mathrm{\mu SFR(UV)} = 51^{+8}_{-6}$ \myr\ from the total UV luminosity ($L_{\rm UV}=1.5\nu L_{\nu}$ with $\nu=\nu(\lambda_{\rm rest}=2800\AA)$) following \cite{bell_2005}. We note that no dust correction was applied to these estimates. We further computed a total SFR estimate (i.e., corrected for the dust attenuation) of $\rm \mu SFR(FUV_{corr})=135^{+171}_{-48}$ \myr\ from FUV luminosities following \cite{hao_2011}, in which the correction to SFR(FUV)$_{\rm obs}$ is less than that applied in \cite{murphy_2011}. This value is in close agreement with $\rm \mu SFR(H\alpha_{corr})=141^{+42}_{-14}$ \myr\ using the same set of calibrations ($\mathrm{log(SFR(H\alpha)} / M_\odot\, \mathrm{yr}^{-1}) = \mathrm{log}(L(\mathrm{H}\alpha) / \mathrm{erg\, s^{-1}}) - 41.27$, \citealt{murphy_2011}), but including a $2.5\times$ correction for the $Z\sim25$\% $Z_\odot$ metallicity measured for z6.3 \citep{fujimoto_2024}. In fact, these calibrations are valid under the assumptions of a constant SFH over 100 Myr timescales and solar metallicity. On the one hand, a correction for the metallicity will decrease the H$\alpha$-based SFR by $\sim0.3-0.4$ dex (a factor of $\sim2-2.5$) because of higher production efficiency of ionizing photons by lower-metallicity, massive, and binary stellar systems (the exact conversion depends on the assumed models, e.g., \citealt{kennicutt-evans_2012, theios_2019, shapley_2023}). The metallicity effect only mildly affects the FUV output \citep{kennicutt-evans_2012}. The IR emission should follow that in the FUV, but the dust opacity decreases at low metallicity, reducing the IR emission at fixed SFR. The total SFR(\ha$_{\rm corr}$) value is broadly consistent with the sum of the IR and (uncorrected) FUV-based SFR estimate (Table \ref{tab:measurements}). These total values are in slight excess, but consistent with the that in F21. This also confirms that this galaxy is consistent with the locus of the main-sequence in the \mstar-SFR plane at $z\sim 6$, as argued in F21.

\subsection{Fraction of obscured SFR}
\label{sec:obscured_sfr_fraction}
Combining these estimates, we derive the obscured SFR fraction for z6.3 of $f_{\rm IR} = \mathrm{SFR(IR) / (SFR(IR)+SFR(FUV))} = 0.76^{+0.14}_{-0.18}$. Based on the upper limit on \lir(\coseven), this fraction decreases to $f_{\rm IR}<0.52$. These values are in agreement with the individual IR-detections for $z\sim7$ UV-selected sources in \cite{algera_2023} at $M_\star\gtrsim10^{9}$ \msun\ (see also \citealt{inami_2022, mitsuhashi_2023_serenade}). Our more constraining upper limit on \lir(\coseven) better agrees with their $f_{\rm IR}$ from stacking of the whole sample at similar \mstar, including non-IR detections. Similar values are reported also for more massive galaxies on the main-sequence at $z\sim5$ \citep{fudamoto_2020, mitsuhashi_2023_cristal} and are in overall agreement with the $f_{\rm IR}$ estimates for mass-complete samples at $z<2.5$ \citep{whitaker_2017}. This is also consistent with the absence of strong \lya\ emission from the lensed images (rest-frame equivalent width (\lya) $\lesssim4$ \AA\ for both z6.3 and the arc at $3\sigma$, F21). Nonetheless, as mentioned above, it is useful to remember that our target was serendipitously discovered as a strong \cii\ and dust continuum emitter in the first place and not preselected based on its UV emission or some rest-frame UV or optical line emission (which might miss dusty objects) as for the samples in \cite{algera_2023} and \cite{fudamoto_2020}. Therefore, it does not suffer from the same selection biases of these literature works and might be a suitable representative of the more dust-rich population of \subl\ galaxies at these redshifts.
\begin{figure}
    \centering
    \includegraphics[trim={0cm 0cm 2cm 2cm}, width=\columnwidth]{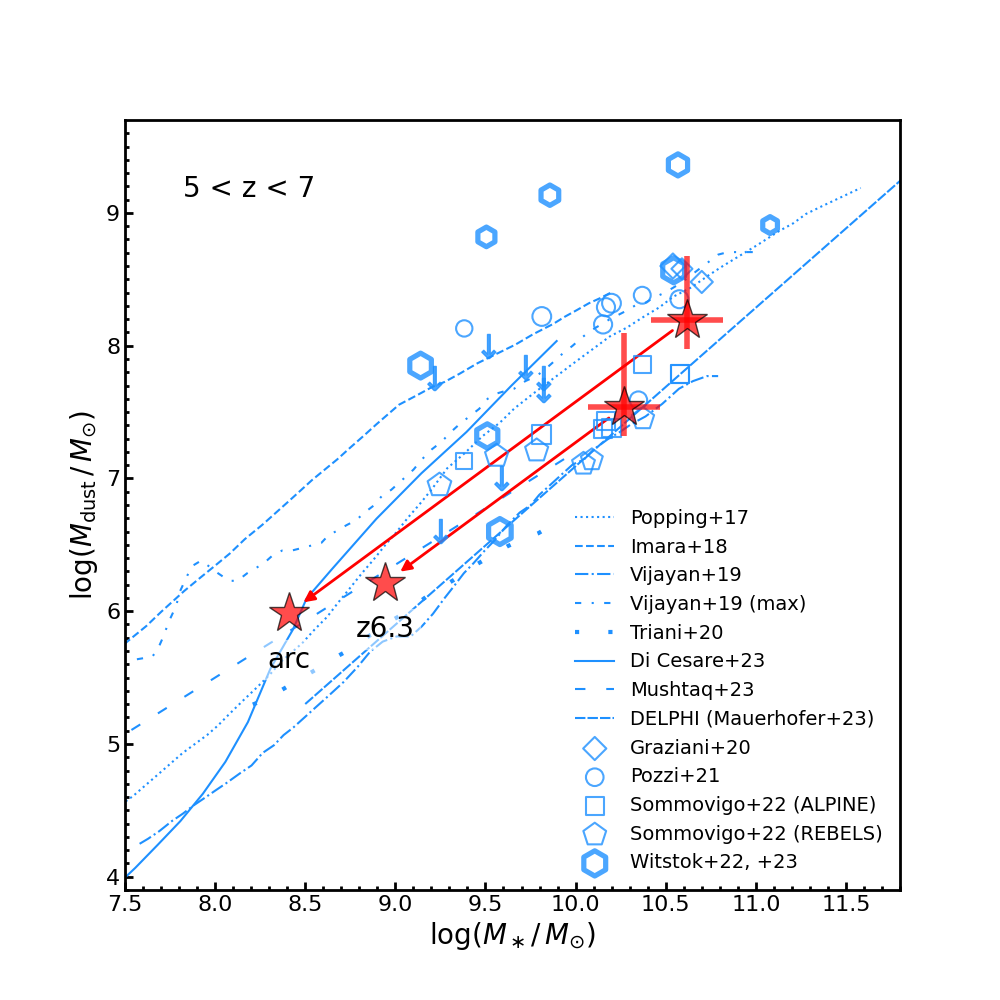}
    \caption{Dust mass as a function of stellar mass at $5<z<7$. The red stars indicate the arc and z6.3 images of our galaxy, as labeled. Red arrows show the direction of the correction for the lensing magnification. The open blue symbols mark observed galaxies from the works in the legend.\ The empirical estimates are from \cite{graziani_2020}, \cite{pozzi_2021}, \cite{sommovigo_2021, sommovigo_2022}, and \cite{witstok_2022,witstok_2023}; the models are from \cite{popping_2017}, \cite{imara_2018}, \cite{vijayan_2019}, \cite{triani_2020}, \cite{di-cesare_2023}, \cite{mushtaq_2023}, \cite{dayal_2022}, and \cite{mauerhofer_2023}. Arrows indicate upper limits. Blue lines show the predictions for the models listed in the legend.}
    \label{fig:dust_production}
\end{figure}

\subsection{Dust production}
\label{sec:dust_production}
The direct dust detection with ALMA and the fraction of obscured star formation point at efficient dust formation in this \subl\ galaxy at $z=6$. The $f_{\rm dust}=M_{\rm dust}/M_{\star}$ fraction can offer an indication of the necessary dust yields. Considering the young average age of the galaxy ($100-200$ Myr, \citealt{gimenez-arteaga_2024}), the dust production via SNe is a more plausible channel than via longer-timescale AGB stars. Following the recipe for a \cite{chabrier_2003} IMF presented in \cite{michalowski_2015}, the galaxy-integrated $f_{\rm dust}=0.002^{+0.003}_{-0.001}$ from z6.3 is converted into a yield of $y_{\rm SNe}=f_{\rm dust}\times 84 = 0.19^{+0.24}_{-0.10}$ \msun.
This yield is in agreement with that reported in L21 ($y_{\rm SNe}=0.3^{+0.5}_{-0.2}$ \msun). The magnified \mdust\ and \mstar\ are consistent with observed values in the local Universe and with the locus of efficient SNe dust production computed by \cite{di-cesare_2023} and \cite{witstok_2023} under similar assumptions as those adopted here.\\

In Figure \ref{fig:dust_production}, we draw a comparison with more complex dust formation models and observations of objects at $z=5-7$ available in the literature \citep[][and references therein]{di-cesare_2023}, complemented with a few more recent works. We corrected the stellar mass estimates of the observations and models derived under the assumption of a \cite{salpeter_1955} IMF to the prescriptions in \cite{chabrier_2003} or \cite{kroupa_2001}, which are similar. We did not attempt to correct \mdust\ for the different IMFs (e.g., influencing the rate of SNe) in the models, given the complex relations among parameters. The choice of different IMFs, along with SFHs, stellar population synthesis models, and several other assumptions in the SED modeling, does remain a major source of systematic uncertainties that inflates the scatter of the distribution in Figure \ref{fig:dust_production}. We highlight the reanalysis of high-redshift sources in \cite{witstok_2022,witstok_2023} as we used a consistent approach, code, and priors to model the FIR emission, and thus a direct comparison can be drawn in terms of \mdust. Both the global and localized measurements in the periphery of the galaxy from the z6.3 and arc images, respectively, are broadly consistent with the existing models and can be explained by SNe without invoking the contribution of AGB or efficient grain growth. Nevertheless, a contribution from growth in the ISM to the dust mass budget is not excluded if we reasonably consider destruction processes such as SN shocks, astration, or ejection (see Figure 9 in \citealt{di-cesare_2023} for a depiction of the dust mass budget for combination of processes at work at high redshift).\\ 

Finally, we cannot avoid noting that we are at a stage where systematic errors totally dominate the uncertainties on \mdust\ and \mstar\ from observations. Even with a common set of observations, the derived physical quantities may vary by up to 1 dex. This currently hinders a critical test of theoretical models and simulations -- the latter predicting \mstar/\mdust\ ratios that vary by up to one order of magnitude at $M_\star \lesssim 10^{9}\,M_\odot$. 

\subsection{The cosmological context}
\begin{figure}
    \centering
    \includegraphics[width=\columnwidth]{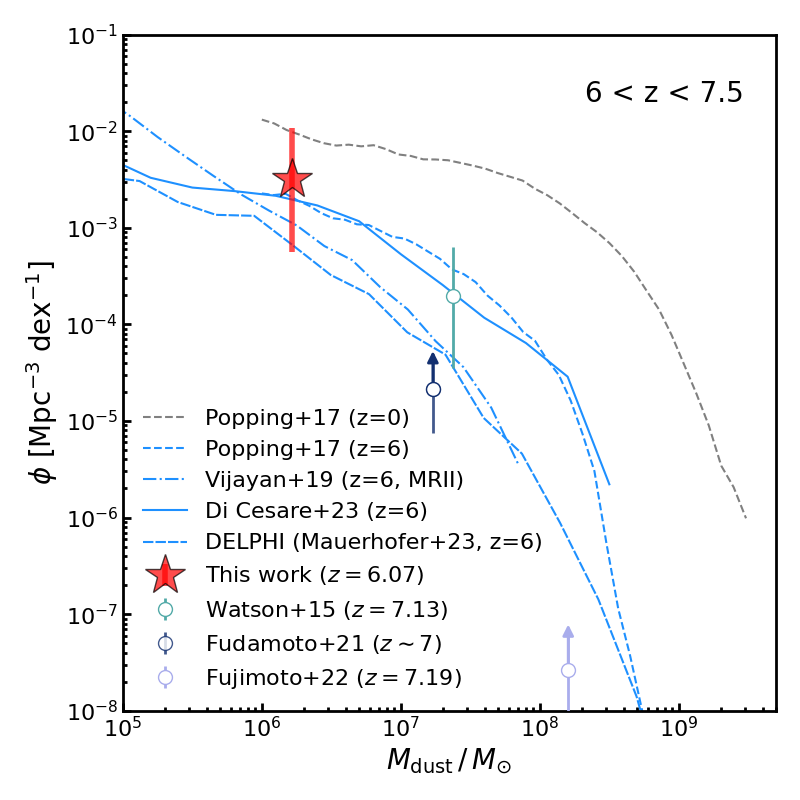}
    \caption{Dust mass function. The blue lines show the predictions of the models at $z=6,$ as labeled (\citealt{popping_2017, vijayan_2019, di-cesare_2023}; \textsc{Delphi}, \citealt{dayal_2022, mauerhofer_2023}). For reference, the dashed gray  line marks the dust mass function from the fiducial model in \cite{popping_2017} at $z=0$. The filled red star and open circles indicate the observational constraints based on our target and other surveys at high redshifts, as labeled (\citealt{watson_2015, bakx_2021, akins_2022, fudamoto_2021, fujimoto_2022}; $\phi$ values estimated in \citealt{fujimoto_2023_counts}).} 
    \label{fig:dmf}
\end{figure}

Given the promising increase in new ALMA detections of intrinsically low-mass and faint UV galaxies at $z\sim 6$ \citep{glazer_2023,fudamoto_2024}, we attempted to place our results in a broader cosmological framework. F21 and \cite{fujimoto_2023_counts} already discussed how our target compares with brighter IR emitters in context of 1.1 mm number counts and the derived \lir\ luminosity function. Here we took advantage of the better constraints on \mdust\ enabled by the multiple ALMA detections in the Rayleigh-Jeans tail of the SED. Figure \ref{fig:dmf} shows the constraint on the dust mass function at $6 \lesssim z \lesssim 7.5$ that we derived from our measurement at intrinsic low masses (i.e., after applying the magnification correction to z6.3). The number statistics are identical to those in \cite{fujimoto_2023_counts} to facilitate the comparison, where the volume has been corrected for the lensing effect and the Poissonian error bars are computed following \cite{gehrels_1986}. For reference, we show the results for the serendipitous detections in the REBELS survey at $z\sim7$ \citep{fudamoto_2021, bouwens_2022}, where we considered the published estimates of \mdust, comoving volume, and correction for clustering. We opted to show this as a lower limit, considering the uncertainties on its derivation (\citealt{algera_2023} for a comparison among different estimates on the SFR density from a similar set of observations from REBELS). Still following Section 6.1 in \cite{fujimoto_2023_counts}, we added constraints from sources in cosmological fields and with the Rayleigh-Jeans tail of the SED sampled by two or more bands. These include a source at $z=7.13$ (\citealt{watson_2015}; we adopt \mdust\ from \citealt{bakx_2021} and increased it by $\sim40$\% to account for the extended emission as in \citealt{akins_2022}) and a dust-enshrouded quasar at $z=7.19$ \citep{fujimoto_2022}. Finally, we show the dust mass functions from a handful of theoretical models (\citealt{popping_2017, vijayan_2019, di-cesare_2023}; \textsc{Delphi}, \citealt{dayal_2022, mauerhofer_2023}). For thoroughness, within the supplementary material, we present an alternative perspective of this figure through cumulative number counts\footnote{Available on Zenodo doi: \href{https://doi.org/10.5281/zenodo.10703293}{10.5281/zenodo.10703293}}.

Considering all the caveats and uncertainties discussed throughout this work and the low number statistics (dominant over the exact choice of the log-bin size in the case of single sources), we find that the models shown in Figure \ref{fig:dmf} are in the right ballpark of the constraints set by our blind survey at $z=6$ and targeted surveys at slightly higher redshifts. A significant drop of the number density of low-\mdust\ galaxy is expected beyond the peak of the cosmic SFR density. For reference, in Figure \ref{fig:dmf} we show the evolution of the dust mass function from $z=0$ to $6$ ($\sim12.5$ Gyr) predicted by the fiducial model by \cite{popping_2017}, able to broadly reproduce the observational estimates at $z=0$ up to \mdust\ $\sim10^{8.3}$ \msun\ \citep[e.g.,][]{vlahakis_2005, dunne_2011, clemens_2013, pozzi_2020}. A decrease by a factor of $\sim10$ in the number density of \mdust\ $\sim10^{6}$ \msun\ galaxies is expected, consistent with the predictions from alternative models and with our fiducial estimate at face value. Our constraint, being formally consistent with scattered observational estimates at $z=0$ (Figure 7 in \citealt{popping_2017}, \citealt{pozzi_2021}), is currently not sufficient to pinpoint any redshift evolution (or shape of the mass function). However, it is a first step in the right direction: forthcoming blind surveys and detailed multiband follow-up to determine \mdust\ will populate Figure \ref{fig:dmf} and inform us on the global dust formation in a cosmological context. 

\section{Conclusions}
\label{sec:conclusions}
We have presented a detailed physical characterization of the global properties of a \subl\ lensed galaxy at $z=6$ with particular emphasis on the cold dust and gas components. Dust and gas were traced in a 60-hour ALMA campaign that covered the emission from $\sim90$ to $\sim370\,\mu$m rest-frame wavelengths and several gas-tracing emission lines. Based on these and ample ancillary data, including JWST imaging and integral field spectroscopy:
\begin{enumerate}
    \item We derive a total SFR (unobscured from FUV and \ha\ lines, obscured from the FIR SED modeling and with an upper limit on \coseven) of $\mu\mathrm{SFR}\sim140$ \myr\ (lensing-corrected to $\sim7$ \myr). Coupled with an updated estimate of $\mu M_\star\sim1.5\times10^{10}$\msun\ ($M_\star\sim7.5\times10^{8}$ \msun\ de-lensed), this source falls on the locus of main-sequence galaxies, while being intrinsically $\sim3$  times fainter than the knee of the luminosity function at $z=6$. 
    \item We detect the rest-frame FIR continuum emission at six different wavelengths with ALMA, robustly confirming the presence of dust and obscured star formation at low stellar masses and metallicities ($Z\sim25$\% \zsun\ from \jwst/NIRSpec). 
    \item We estimate dust temperatures in the range \tdust$\sim37-50$K, the lower end being observed in a highly magnified peripheral region of the galaxy. This might suggest the presence of a negative \tdust\ gradient, but the resolution, S/N, and lack of detections close to the peak of the dust emission currently hamper our ability to constrain \tdust\ and its possible spatial variations.
    \item We constrain the amount of obscured star formation to $f_{\rm IR}\lesssim50-70$\%, which is consistent with trends at lower redshifts and a handful of detections and stacking at $5 < z < 7$. The main uncertainties in this calculation are on \tdust\ and, by extension, on the SFR(IR).   
    \item We estimate a dust mass of $\mu M_{\rm dust}\sim0.3\times10^{8}$\msun\  ($M_{\rm dust}\sim1.5\times10^{6}$ \msun\ de-lensed) and a dust-to-stellar mass fraction of $f_{\rm dust}\sim0.002$. This amount of dust is consistent with a rapid production from SNe without invoking AGB stars, which are unlikely to contribute due to the young age of the galaxy. Our measurements are broadly consistent with the wide range of predictions from more complex models of dust evolution.  
    \item We estimate a total gas mass, or upper limits, based on \citwo, \cii, and dynamical modeling. The dynamical modeling sets an intrinsic maximum amount of $M_{\rm gas}\sim 2\times10^9$ \msun, consistent with the upper limit on \citwo\ and some of the calibrations of the \cii-\mgas\ relation. This implies a dust-to-gas mass ratio, $\delta_{\rm DGR}$, on the order of $\sim10^{-3}$, in good agreement with models and empirical $Z-\delta_{\rm DGR}$ relations in the literature.
    \item Armed with an estimate of \mdust\ more robust and lower than typical values in the literature at these redshifts, we placed a constraint on the dust mass function at the low-\mdust\ end. Encouragingly, we find the predictions of a few recent dust formation models to be in the ballpark of our observational constraint despite all the caveats and uncertainties.
\end{enumerate}

This is an example of how the synergy among cutting-edge instruments collecting deep observations across the electromagnetic spectrum opens a window onto the physics of numerous low-luminosity, low-mass, low-SFR, and low-metallicity galaxies at reionization. Future works in preparation and planned targeted observations (e.g., a ALMA high-frequency follow-up to pin down \tdust\ or at higher angular resolutions) will allow us to explore the cold gas and dust properties on sub-galactic scales. Larger samples will eventually be assembled and will be the key to constraining correlations and models, including the global production of dust in a cosmological framework, a task currently impossible given their scatter and the low number of galaxies with data of the necessary quality and coverage.

\begin{acknowledgements}
We thank the referee for their insightful comments that improved this article. We are grateful to Aswin Vijayan, Pratika Dayal, and Valentin Mauerhofer for sharing their models and predictions. FV warmly thanks Joris Witstok for help and support in using the fitting code \textsc{Mercurius}; Kasper Heintz, Gerg\"{o} Popping, and the rest of the GESO team at the European Southern Observatory for useful discussions while preparing this manuscript. The Cosmic Dawn Center (DAWN) is funded by the Danish National Research Foundation under grant No. 140. S.F. acknowledges the support from NASA through the NASA Hubble Fellowship grant HST-HF2-51505.001-A awarded by the Space Telescope Science Institute, which is operated by the Association of Universities for Research in Astronomy, Incorporated, under NASA contract NAS5-26555. FEB acknowledges support from
ANID Millennium Science Initiative Program - ICN12\_009, CATA-BASAL - FB210003, and FONDECYT Regular - 1200495 (FEB); DE acknowledges support from a Beatriz Galindo senior fellowship (BG20/00224) from the Spanish Ministry of Science and Innovation,projects PID2020-114414GB-100 and PID2020-113689GB-I00 financed by MCIN/AEI/10.13039/501100011033, project P20-00334  financed by the Junta de Andaluc\'{i}a, and project A-FQM-510-UGR20 of the FEDER/Junta de Andaluc\'{i}a-Consejer\'{i}a de Transformaci\'{o}n Econ\'{o}mica, Industria, Conocimiento y Universidades. G.E.M. acknowledges the Villum Fonden research grants 13160 and 37440.
This paper makes use of the following ALMA data: ADS/JAO.ALMA\# 2021.0.00055.S, 2021.0.00181.S, 2021.0.00247.S, and 2022.0.00195.S. ALMA is a partnership of ESO (representing its member states), NSF (USA) and NINS (Japan), together with NRC (Canada), MOST and ASIAA (Taiwan), and KASI (Republic of Korea), in cooperation with the Republic of Chile. The Joint ALMA Observatory is operated by ESO, AUI/NRAO and NAOJ. The research is also based in part on observations made with the NASA/ESA/CSA \textit{James Webb} Space Telescope. The data were obtained from the Mikulski Archive for Space Telescopes at the Space Telescope Science Institute, which is operated by the Association of Universities for Research in Astronomy, Inc., under NASA contract NAS 5-03127 for JWST. These observations are associated with program \#1567.
\end{acknowledgements}

\bibliographystyle{aa} 
\bibliography{bib_z6_dust_alcs} 

\begin{appendix}

\section{Tests on the photometric extraction and SED modeling}
\label{app:alternative_photometries}

\begin{landscape}
\begin{figure}
    \centering
    \includegraphics[trim={2cm 0 10cm 0}, scale=0.4]{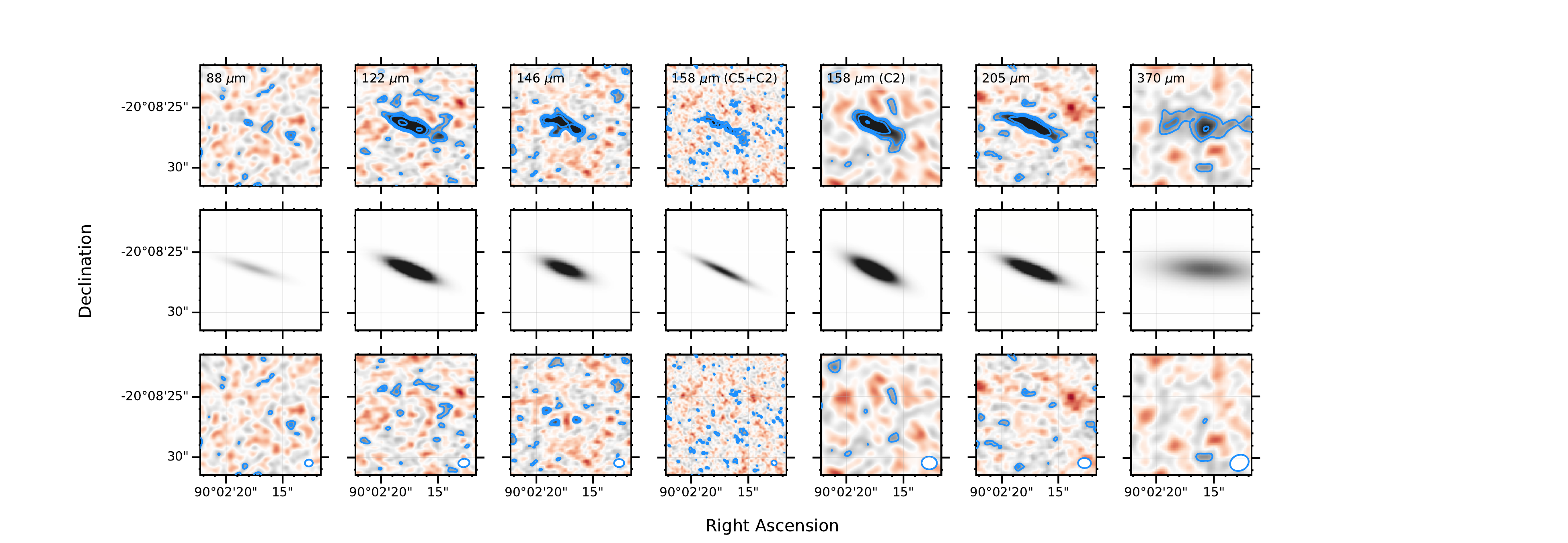}
    \includegraphics[trim={2cm 0 10cm 2cm}, scale=0.4]{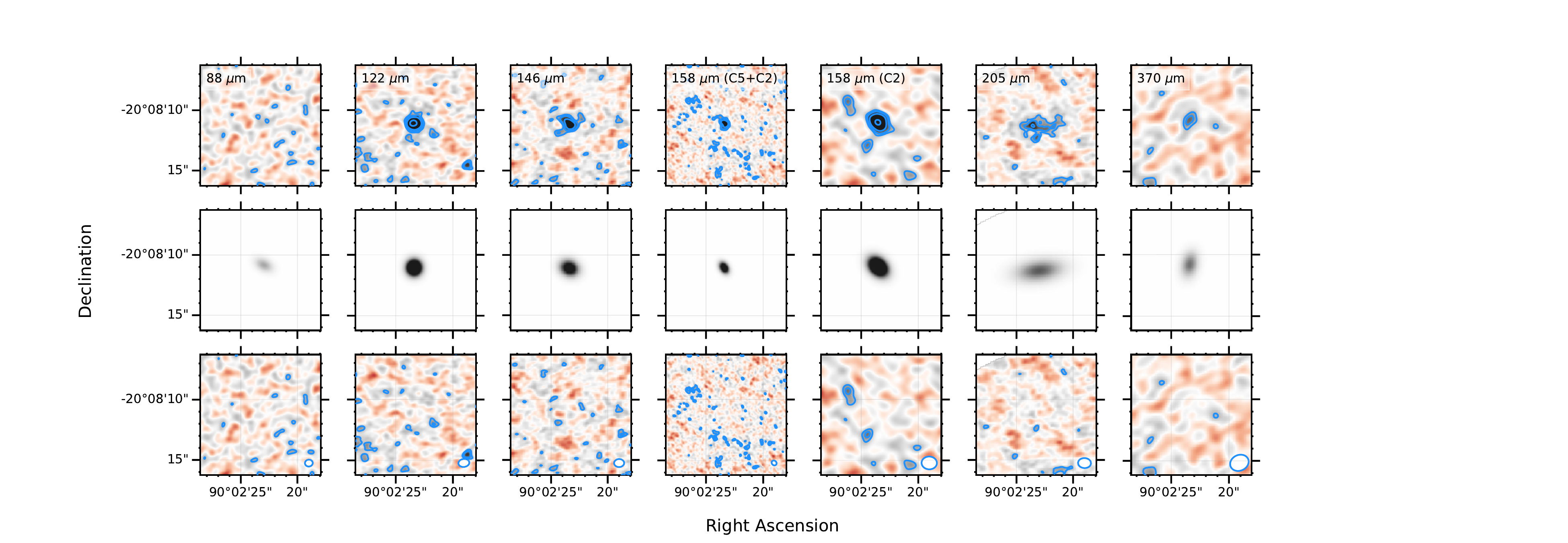}
    \caption{Dust continuum maps, best-fit Gaussian modeling, and residuals for the lensed z6.1-6.2 (arc, top rows) and z6.3 (bottom rows) images. The side of each cutout is 10". The rest-frame wavelengths of each map are labeled. Blue contours are at $2,\,3,\,5,$ and $10\sigma$ (rms). The images are color-scaled within $\pm5$ times the rms per pixel in each band. The beam is shown in the bottom-right corner of the cutouts of the residuals.}
    \label{fig:app:imfit}
\end{figure}
\end{landscape}

\begin{figure*}
    \centering
    \includegraphics[width=\textwidth]{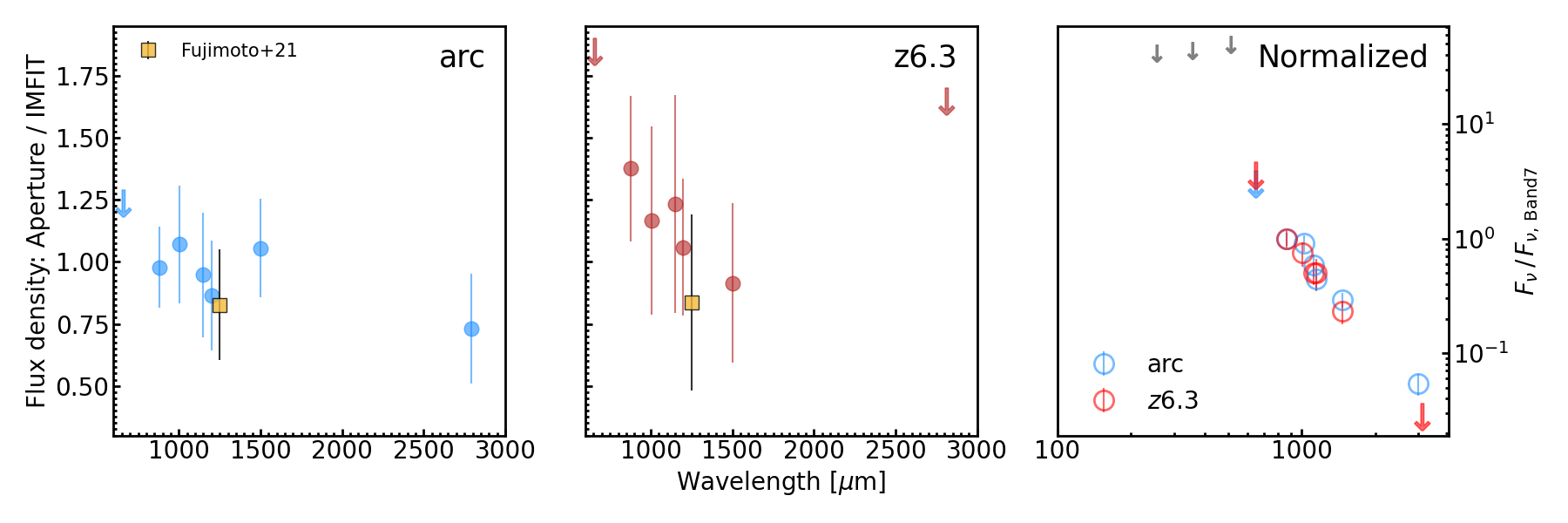}
    \caption{Photometric extractions. Filled circles indicate the ratio between the observed photometry extracted in Kron apertures and that from modeling with \textsc{CASA/IMFIT}. Large arrows show the $2\sigma$ upper limits on aperture photometry. Small gray arrows are $3\sigma$ upper limits on \textit{Herschel}/SPIRE from \cite{sun_2022}. For reference, we show both the measured Band 6 flux densities in configuration C2 and the combined C2+C5 (offset to improve the readability). \textit{Left}: Arc (z6.1-6.2). \textit{Center}: z6.3. \textit{Right:} Aperture photometry of both lensed images. The flux densities are normalized to that in Band 7 ($122\,\mu$m rest frame) for clarity and to avoid introducing the uncertainties on the magnification factor. The measurements in Band 6 ($158\,\mu$m rest frame) from \cite{fujimoto_2021} are shown as orange squares.}
    \label{fig:alternative_photometries}
\end{figure*}

In Figure \ref{fig:app:imfit}, we show the results of the best-fit elliptical Gaussian modeling with \textsc{CASA/IMFIT} (Section \ref{sec:photometry}). For detections of both the arc and the z6.3 image, the total flux density from this modeling is consistent with that from aperture photometry (Figure \ref{fig:alternative_photometries}). This suggests that the clumping evident in Bands observed at higher resolution (Band 6 at $158\,\mu$m and shorter rest-frame wavelengths mapped by \jwst) does not introduce major bias in our global estimates. Also, the detection in at $370\,\mu$m rest frame is less significant than in every other band except at $88\,\mu$m. The elongated western tail in the arc image is consistent with noise.\\

Furthermore, as part of the supplementary material available online, we include variations on modeling of the FIR/millimeter SEDs with an alternative choice for the prior on $\beta$ (flat within $[1,5]$) and adopting the self-consistent treatment to estimate the wavelength at which the optical depth reaches unity \citep[$\lambda_{0}(\tau =1)$,][]{witstok_2023}. We applied this treatment to z6.3 after de-lensing and adopting the intrinsic effective radius in the source plane (Section \ref{sec:fir_sed}). This approach confirms that, on average, the optical thin assumption holds for our target and that the ensuing best-fit parameters are consistent with those listed in Table \ref{tab:best-fit} for a fixed sets of priors. Concerning the choice of the priors on the dust emissivity slope, $\beta$, the best-fit estimates (i.e., the median of the posterior distribution) determined using a uniform distribution are 15\% higher and 20\% lower for z6.3 and the arc, respectively, than using the physically informed Gaussian distribution. The error bars (i.e., the $16-84\%$ inter-percentile range) also generally increase and the final estimates are overall consistent with the values presented in Table \ref{tab:best-fit}. The uncertainties connected with the exact choice on the priors (or fixed values) are known \citep{juvela_2013, bakx_2021, algera_2023} and add up to the error budget and systematics when comparing different works in the literature. All corner plots for the SED models discussed in this work are available online as supplementary figures.

\FloatBarrier

\section{Literature sample}
\label{app:literature}

In Figure \ref{fig:lir_lprime-lir}, we show the measurements for our z=6 \subl\ galaxy in comparison with literature samples with available \coseven, \citwo\ luminosities, and reliable dust continuum detections to derive \lir\ and \mdust. \lir\ refers to the total IR luminosity integrated between 8 and 1000 $\mu$m. Estimates of FIR luminosities integrated over narrower ranges (e.g., 40-122 $\mu$m) have been corrected to total with an average correction factor of $\sim 1.3$. We note that similar corrections might apply depending on the exact models used in the FIR SED modeling (e.g., a modified black body predicts lower \lir\ luminosities than physically motivated \cite{draine_2007} models, \citealt{berta_2016, valentino_2018}). Corrections for lensing were generally not applied. The samples include those already described in \citealt{valentino_2020c}, expanded to more recent results: local galaxies from the \textit{Herschel}-FTS archive \citep{liu_2015, liu_2021} and the HerCULES \citep{rosenberg_2015} survey; main-sequence galaxies at $z\sim1.5$ \citep{valentino_2020b, valentino_2020c}; blindly detected sources from ASPECS \citep{boogaard_2020}; variously selected submillimeter and IR-bright galaxies at $z=2-6$, including quasars \citep{walter_2011, strandet_2017, canameras_2018, nesvadba_2019, yang_2017, andreani_2018, cortzen_2020, riechers_2020, harrington_2021, gururajan_2022, decarli_2022, lei_2023, gururajan_2023, hagimoto_2023}.

\end{appendix}

\end{document}